\documentclass[fleqn,usenatbib]{rasti}
\usepackage{newtxtext,newtxmath}
\usepackage[T1]{fontenc}
\DeclareRobustCommand{\VAN}[3]{#2}
\let\VANthebibliography\thebibliography
\def\thebibliography{\DeclareRobustCommand{\VAN}[3]{##3}\VANthebibliography}  

\usepackage{graphicx}
\usepackage{amsmath}
\usepackage{tabularx}
\usepackage[english]{babel}
\usepackage[utf8]{inputenc}
\usepackage{array}
\usepackage{siunitx}
\usepackage{comment}
\usepackage[switch, modulo]{lineno}         
\usepackage{caption}
\usepackage{placeins}
\usepackage{threeparttable}
\usepackage{booktabs}
\usepackage{bm}   

\providecommand{\tens}[1]{\mathsf{#1}} 
\renewcommand{\vec}[1]{\bm{#1}}  
\usepackage{needspace}
\usepackage{etoolbox}

\title{A guide to choosing data compression methods for cosmological inference}
\author[S. Pyne et al.]{
Susan Pyne,$^{1}$\thanks{E-mail: s.pyne@ucl.ac.uk}
Benjamin Joachimi,$^{1}$
Calum Gordon$^{1}$
and Kiyam Lin$^{1}$\\
$^{1}$Department of Physics \& Astronomy, University College London, Gower Street, London, WC1E
6BT, UK
}
\date{Accepted XXX. Received YYY; in original form ZZZ}

\pubyear{\the\year{}}

\begin{document}
\label{firstpage}
\pagerange{\pageref{firstpage}--\pageref{lastpage}}
\maketitle    

\begin{abstract}

  We provide a pedagogical guide to help researchers choose an appropriate compression method for cosmological inference and numerical covariance estimation, trading off the balance between information loss and complexity.  We describe methods used in the literature, categorising them by the form of the loss function -- Fisher information, mutual information, or mean squared error -- and by whether they are linear or non-linear. 
 We consider in detail the linear compression methods: massively optimised parameter estimation and data compression (MOPED), canonical correlation analysis (CCA), principal component analysis, and linear neural networks. We also investigate simple non-linear methods based on neural networks with similar architectures but different loss functions. We use tomographic weak lensing power spectra as example data vectors, deriving constraints on the cosmological parameters $\Omega_\mathrm{m}$, $\sigma_8$, $w_0$ and $w_a$ from compressed data.  For our figure of merit (FoM) we use the log determinant of the Fisher matrix, which approximates the covariance of the full posterior well.  MOPED and a Fisher information-based linear neural network both achieve lossless compression. Neural networks with loss functions which optimise mutual information achieve at most 84 per cent of the uncompressed FoM and those which minimise mean squared errors achieve at most 72 per cent.  We show how to determine whether MOPED will retain information even if the fiducial cosmology is not known with certainty, and demonstrate a strategy for CCA which results in an FoM close to that of MOPED while requiring only an approximate idea of where the bulk of the posterior mass is located.\\
\end{abstract}


\section{Introduction}\label{sect:intro}

In cosmology it is becoming increasingly common to compress long data vectors which it would otherwise be infeasible to incorporate into inference pipelines or to derive numerical covariance estimates. There are many drivers of the need for compression. Standard two-point analyses are being supplemented by higher-order summary statistics or by field-level analyses which do not use conventional summary statistics at all. In these cases it is generally impossible to formulate an analytical expression for the likelihood.   This has inspired the development of simulation-based inference (SBI) in which the likelihood is not pre-specified, but learned by a neural network \citep{cranmer2020frontier, thiele2026machine}.  SBI nearly always involves a data compression step to allow inference to be carried out in simpler latent spaces which are maximally coupled to the parameters (for example \citealt{alsing2018generalized, jeffrey2021likelihood, lin2023simulation, gatti2024dark,  jeffrey2025dark, williamson2026dark}).  Similarly, current surveys require  very large covariance matrices. Estimating these numerically almost always involves significant data compression \citep{reischke2025kids}. Moreover data and covariances may come with complicated or poorly understood systematic uncertainties involving many nuisance parameters which again point to the need for compression. 

This raises the question of how to compress data in an optimal but practical way.  Compression will usually be only a small element of the analysis pipeline, rather than an end in itself, so the aim would normally be to reduce the length of the data vector in a way which retains as much information as possible but is not unduly resource-intensive or difficult to implement, and does not depend on overly restrictive assumptions.  In the context of simulation-based inference,  \cite{alsing2018generalized} and \cite{jeffrey2021likelihood}, among others, have pointed out that compression which loses some information will not bias the inference results, although it may widen the constraints on parameters. So it may be reasonable to choose a compression method which loses some information if this simplifies the analysis as a whole. In particular an analytic method could be preferred to a complex non-linear neural network solution even if it does not retain as much information.  It is important to make this trade-off explicit.

Several previous works have compared compression methods for cosmology. \cite{sharma2024comparative} compared the compression performance of different loss functions in convolutional neural networks using weak lensing data.  However they combined this with a comparison of network architectures, so their conclusions are not purely about compression.  \cite{lanzieri2025optimal} reviewed and compared neural network-based compression methods for full-field weak lensing and found empirically that an information-theoretical method, variational mutual information maximisation, could be close to lossless. \cite{park2025dimensionality} carried out a comprehensive comparison of linear and non-linear methods, illustrated by two- and three-point weak lensing statistics. They concluded that linear methods could perform just as well as non-linear in some circumstances and provided some practical guidelines. 

The aim of the current work is more pedagogical: to give an overview of the most commonly used compression methods, comment on their advantages  and disadvantages, and highlight practical considerations and questions which may arise in their use, with the aim of helping researchers to choose the best method for their purposes.  This is not intended to be a comprehensive review and we do not try to cover every compression method which has been suggested for cosmology (some recent examples which we exclude are  \citealt{byun2017towards}, \citealt{gualdi2020geomax},  \citealt{philcox2021fewer}, and \citealt{akhmetzhanova2024data}). On the other hand much of our discussion is also applicable to fields outside cosmology.

We illustrate the methods with the weak lensing power spectrum. This is a well-understood and analytically tractable summary statistic, suitable for our pedagogical aims, even though it does not exploit the full non-Gaussian information contained in field-level data or higher-order summary statistics. Our analysis is based on a $w_0w_a$CDM cosmological model, in which we consider only the four cosmological parameters  which weak lensing is most sensitive to, $\Omega_\mathrm{m}$, $\sigma_8$, $w_0$ and $w_a$, and do  not include any nuisance parameters describing systematic effects.  The survey assumptions are similar to \textit{Euclid}\footnote{https://www.euclid-ec.org/}, but with only three tomographic bins.  

Given that we are interested in how much information is retained after compression, in Sect.~\ref{sect:info} we  first  set out two principled definitions of information which have both been used in cosmology. These are  Fisher information, which quantifies the sensitivity of the data to the parameters,  and  mutual information, which quantifies the reduction in uncertainty in one random variable given knowledge of another. 

In Sect.~\ref{sect:FI} we discuss compression methods based on Fisher information: the massively optimised parameter estimation and data compression algorithm  (MOPED) \citep{heavens2000massive}, compression to the score \citep{alsing2018generalized}, and information-maximising neural networks  (IMNN) \citep{charnock2018automatic}. Sect.~\ref{sect:MI} covers commonly used methods with loss functions related to mutual information  or which optimise mutual information in special circumstances.  These are: canonical correlation analysis (CCA),  variational mutual information maximisation   (VMIM)  \citep{jeffrey2021likelihood}' and a special case of this known as Gaussian negative log likelihood minimisation  (GNLL) \citep{fluri2018cosmological}.  Sect.~\ref{sect:other} presents methods which are based on minimising squared errors rather than on a specific definition of information: principal component analysis (PCA), and neural networks which minimise the mean squared error of the predicted parameters.  We also introduce linear versions of the main non-linear methods: neural networks with no non-linear features. These allow us to compare linear neural networks with analytical counterparts.  Table~\ref{tab:compression_methods} summarises the compression methods we consider, indicating whether they are linear or non-linear and whether they involve a neural network.

Section~\ref{sect:tests} describes how we test the performance of the main methods. Section \ref{sect:results} reports our findings and  explores some practical issues in implementing them, in particular MOPED and CCA.  We conclude with recommendations in Sect.~\ref{sect:conclusions}. 
Appendices give further derivations and background, and summarise connections between the methods.
\begin{table*}
    \centering
    \begin{threeparttable}
     \caption{Compression methods discussed in this paper}
    \label{tab:compression_methods}
    \begin{tabular}{llcccc}
    \toprule
    
   & &  {\textbf{Linear/}} &\textbf{Neural} & \textbf{Described in}&\textbf{Results in}\\
      \multicolumn{1}{c}{ \textbf{Method}}&\textbf{Designation}&\textbf{non-linear}&\textbf{network?}&\textbf{section}&\textbf{section}\\
    
    \midrule
    \textbf{Fisher information-based loss function}\\
     Massively optimised parameter estimation and data compression &    MOPED &L&no& \ref{sect:MOPED}&\ref{sect:FI_results}\\
         Score compression&$-$&N&no&\ref{sect:score}&$-$\\ 
         Linear neural network which maximises FI&LNN-FI &L&yes&\ref{sect:LNN-FI}&\ref{sect:FI_results}\\
         Non-linear neural network  which maximises FI & NN-FI  &N&yes&\ref{sect:LNN-FI}&\ref{sect:FI_results}\\
         
         Information-maximising neural network&IMNN&N&yes&\ref{sect:IMNN}&$-$\\
    \\
    \textbf{Mutual information-based loss function}\\
          Canonical correlation analysis \tnote{1} &CCA & L & no&\ref{sect:CCA}&\ref{sect:MI_results} \\
          Variational mutual information maximization&VMIM  &N&yes& \ref{sect:VMIM}&$-$\\
          Linear neural network with Gaussian negative log likelihood loss\tnote{2}&LNN-GNLL&L& yes&\ref{sect:GNLL}&\ref{sect:MI_results}\\
           Non-linear neural network with Gaussian negative log likelihood loss\tnote{2}&NN-GNLL &N&yes&\ref{sect:GNLL}&\ref{sect:MI_results} \\
          
    \\
    \textbf{Mean squared error-based loss function}\\
         Principal component analysis\tnote{3}&PCA & L&no&\ref{sect:PCA}&\ref{sect:MSE_results}\\
        
        Linear neural network which minimises MSE of parameters&LNN-MSE&L&yes&\ref{sect:NN-MSE}&\ref{sect:MSE_results}\\ 
        Non-linear neural network which minimises MSE of parameters&NN-MSE& N&yes&\ref{sect:NN-MSE}&\ref{sect:MSE_results}\\ 
         
    \bottomrule
    \end{tabular}   
    \begin{tablenotes}
\item[1] Maximises MI only if data is Gaussian
\item[2] Maximises MI under appropriate assumptions
\item[3] Minimises data variance

\end{tablenotes}
\end{threeparttable}
\end{table*} 

\section{What is \lq information\rq?}\label{sect:info}
In Bayesian inference we use data to infer the values of parameters, aiming to obtain the tightest possible posterior distributions: those which are as informative as possible about the parameters.  So to demonstrate that the inference is close to optimal we need a rigorous quantified definition of information which can be used as an optimisation criterion.  In most cosmological analyses to date, information has been measured by the Fisher information matrix which measures the curvature of the log likelihood close to the maximum likelihood point.  A second possible definition is mutual information, a concept from information theory which measures how much information is shared by two variables: in our case by the compressed data and the parameters. This is a more global definition connected to the idea of statistical sufficiency across the whole parameter space \citep{sui2026evaluate}.

\subsection{Fisher information (FI)}
The Fisher information matrix is defined as 
\begin{align}
\tens{F}_{ij} &= - \mathbb{E}\bigg [ \frac{\partial^2 \ln \mathcal{L}}{\partial\theta_i\partial\theta_j}\bigg]\label{eq:FIM}
\ ,
\end{align}
where $\mathcal{L}$ is the log likelihood, $\vec{\theta}$ is the vector of parameters, and the derivatives are evaluated at a fiducial point, ideally close to the maximum likelihood.
 So to linear order the Fisher information is the variance of the score, which is defined as the derivative of the log likelihood with respect to the parameters.
 
 In terms of the data $\vec{d}$ and its covariance $\tens{C}$ the Fisher matrix can be written as 
 \begin{align}
     \tens{F}_{ij} &= \frac{1}{2} \mathrm{Tr}\bigg[\tens{C}^{-1}\frac{\partial\tens{C}}{\partial\theta_i}\tens{C}^{-1}\frac{\partial\tens{C}}{\partial\theta_j} + \tens{C}^{-1}\bigg(\frac{\partial\vec{d}}{\partial\theta_i}\frac{\partial\vec{d}^\intercal}{\partial\theta_j} +\frac{\partial\vec{d}}{\partial\theta_j}\frac{\partial\vec{d}^\intercal}{\partial\theta_i} \bigg)\bigg] \ .
 \end{align}

The Fisher matrix measures the sensitivity of the data to the parameters at the fiducial point. It can be shown that its inverse is a lower bound to the variance of any unbiased estimator of the parameters.  This is known as the Cramér-Rao bound (or alternatively as the information inequality) and is fundamental to the insight that using Fisher information as a loss function can optimise information retention locally.  

\subsection{Mutual information (MI)}
Mutual information is a concept from information theory which quantifies how much the uncertainty about one random variable is reduced by knowing another. In our case the random variables are the parameters and the data.

To define mutual information we start from the information entropy, 
 $H(\vec{x})$, of a probability distribution $P(\vec{x})$. This is defined as
\begin{align}
    H(\vec{x}) &= -\int  \mathrm{d}\vec{x}  P(\vec{x}) \log (P(\vec{x}))  \\   
    &=-\mathbb{E}[\log (P(\vec{x}))] \ . \label{eq:Hx}
\end{align}
The $\log$ function can be to any base. Often base 2 is used because this corresponds to units of bits. In this work we use $\log$ for a general base and $\ln$ when the natural logarithm is specifically intended.

The mutual information $I(\vec{x};\vec{y})$ of two random variables $\vec{x}$ and $\vec{y}$ is then defined as 
\begin{align}\label{eq:MI}
    I(\vec{x};\vec{y}) &= H(\vec{x}) + H(\vec{y}) - H(\vec{x},\vec{y}) \ ,
\end{align}
where $H(\vec{x},\vec{y})$ is the joint entropy, defined as  
\begin{align}
H(\vec{x},\vec{y})&=-\iint \mathrm{d}\vec{x} \mathrm{d}\vec{y}P(\vec{x},\vec{y})\log P(\vec{x},\vec{y}) \ .
\end{align}
Alternatively the mutual information can be expressed in terms of the conditional entropy $H(\vec{y}|\vec{x})$ as
\begin{align}
    I(\vec{x};\vec{y}) &= H(\vec{y}) - H(\vec{y}|\vec{x}) \\
    &= H(\vec{y}) - \mathbb{E}\bigg[\log \frac{P(\vec{x},\vec{y})}{P(\vec{x})}\bigg]\\
    &= H(\vec{y})+\mathbb{E}[\log P(\vec{y}|\vec{x})]\label{eq:MIcond} \ .
\end{align}

Mutual information can also be interpreted as the Kullback--Leibler (KL) divergence between the joint distribution $P(\vec{x},\vec{y})$ and the marginal distributions $P(\vec{x})$ and $P(\vec{y})$. From Eq.~(\ref{eq:MI}) 
\begin{align}
    I(\vec{x}; \vec{y}) &=   - \int   \mathrm{d}\vec{x} P(\vec{x})\log P(\vec{x}) 
    -\int \mathrm{d}\vec{y}P(\vec{y})\log P(\vec{y}) \notag\\
    & \hspace{2cm} +\iint \mathrm{d}\vec{x} \mathrm{d}\vec{y}P(\vec{x},\vec{y})\log P(\vec{x},\vec{y})\\
    &= \iint  \mathrm{d}\vec{x}  \mathrm{d}\vec{y}\ [-P(\vec{x},\vec{y}) \log P(\vec{x}) \notag\\ 
    &\hspace{2cm} - P(\vec{x},\vec{y}) \log P(\vec{y})  + P(\vec{x},\vec{y})\log P(\vec{x},\vec{y}) ]\\
    &= \iint  \mathrm{d}\vec{x} \mathrm{d}\vec{y}\ P(\vec{x},\vec{y})[ -\log P(\vec{x})
    - \log P(\vec{y})\notag\\
    &\hspace{2cm}+\log P(\vec{x},\vec{y})]\\
    &= \iint  \mathrm{d}\vec{x} \mathrm{d}\vec{y} \ P(\vec{x},\vec{y})\log \Bigg[\frac{P(\vec{x},\vec{y})}{P(\vec{x})P(\vec{y})} \Bigg]\ .\label{eq:KLdiv}
\end{align}
The final line is the definition of the Kullback--Leibler  divergence between $P(\vec{x},\vec{y})$ and $P(\vec{x})P(\vec{y})$. 

As a consequence mutual information can be viewed in several equivalent ways:
\begin{itemize}
    \item as the amount of information about one variable obtained by observing a second variable i.e.  their relative entropy;
    \item  as the KL divergence (or \lq  distance\rq ) between the joint distribution of two random variables and the product of their marginal distributions;
    \item  in the context of Bayesian inference, the mutual information between the parameters and data is the expected reduction in uncertainty in moving from the prior to the posterior.
\end{itemize}

\subsection{Which information definition should be used?}
Both concepts of information are useful but they serve different purposes.  Fisher information is a local metric describing parameter sensitivity close to a fiducial point in parameter space whereas mutual information is used to compare the whole posterior across all parameter space. Intuitively Fisher information is more relevant if the posterior is known to be sharply peaked around a single maximum likelihood point whereas mutual information may be more appropriate for a flatter or multi-modal  distribution.
Appendix~\ref{sect:MI_FI} sets out one way of seeing that these two views of information are connected.  Other interesting connections are explored in appendix~B of \cite{sui2026evaluate}.

\section{Compression methods}\label{sect:methods}

Table~\ref{tab:compression_methods} shows the methods which we cover in this work, categorised by the type of loss function used, whether they are linear or non-linear, and whether they involve a neural network. This table also indicates the section of the paper where each is discussed: firstly methods based on Fisher information, then those based on mutual information, and finally  methods with mean squared error loss functions. Where relevant the section where results are given is also shown.
Appendix~\ref{sect:relations} summarises connections between the methods.

\subsection{Compression based on Fisher information}\label{sect:FI}
The seminal investigation of compression methods based on FI is  \cite{tegmark1997karhunen}. These authors explored the problem of finding projections (or equivalently compressions) of the data which locally maximise sensitivity to the parameters, referring to this as  \lq  Karhunen--Lo\`eve\rq.\footnote{ This nomenclature can be confusing. Appendix \ref{sect:Karhunen} gives more background about its history.} They considered the estimation of a single parameter in the special cases where either the mean or the covariance of the data are known, i.e. do not depend on the parameters, and where the likelihood is Gaussian, and then extended the analysis to multiple parameters.  
They showed that if the covariance is known and the likelihood is Gaussian, then the data can be compressed down to a single number, but if the covariance is parameter-dependent then compression is lossy even for a Gaussian likelihood, so quadratic terms are needed. If the likelihood is not Gaussian then at least quadratic and possibly higher order terms are needed, even if the covariance does not depend on the parameters. 

All compression methods based on FI have the disadvantages that they require knowledge of a suitable fiducial point and also require derivatives at that point. In practical situations these can be quite restrictive requirements.

\subsubsection{Massively optimised parameter estimation and data compression (MOPED)}\label{sect:MOPED}

MOPED develops the concepts in \cite{tegmark1997karhunen} to prescribe how to implement lossless compression for multiple parameters in the restricted case where the likelihood is Gaussian and the covariance does not depend on the parameters.  
The full methodology is set out in \cite{heavens2000massive} who showed that it is possible to construct $M$ numbers \mbox{$y_m= \vec{b}_m^\intercal\vec{d}$}  which are  mutually uncorrelated \mbox{($\vec{b}_i^\intercal\tens{C} \vec{b}_j = 0$} for $i\ne j$), such that $y_m$ encapsulates all information about the parameter $\theta_m$ which is not already contained in $y_q$, $q<m$. The vectors 
$\vec{b}_m$ weight the data points according to how informative they are about the parameters $\theta_m$. These weighting vectors are orthogonal and are given by:
\begin{align}
\vec{b}_1 &= \frac{\tens{C}^{-1}\vec{d}_{,1}}{\sqrt{\vec{d}_{,1}^\intercal \tens{C}^{-1}\vec{d}_{,1}}}\\
\vec{b}_m &= \frac{{\tens{C}^{-1}\vec{d}_{,m} - \sum_{q=1}^{m-1}(\vec{d}_{,m}^\intercal
\vec{b}_q)\vec{b}_q }}{
\sqrt{\vec{d}_{,m}^\intercal \tens{C}^{-1}\vec{d}_{,m} - \sum_{q=1}^{m-1}
(\vec{d}_{,m}^\intercal \vec{b}_q)^2}} \ ,
\label{eq:bbm}
\end{align}
where $\vec{d}_{,m}$ is the derivative of the data with respect to the parameter $\theta_m$ at the fiducial point.
The denominator normalises the compression weights and in practice does not affect the compression performance.   Provided that the covariance is independent of the parameters and the likelihood is Gaussian, this compression is lossless: the Fisher information of the original and compressed data can be shown algebraically to be identical (see appendix~A of \citeauthor{heavens2000massive} \citeyear{heavens2000massive}).
\subsubsection{Compression to the score}\label{sect:score}
The score is the derivative of the log likelihood with respect to the parameters. \cite{alsing2018generalized} introduced the idea of using the score as a loss function, generalising the argument in \cite{tegmark1997karhunen} to non-linear models and cases where the likelihood is not Gaussian and is not necessarily known exactly.  

We can Taylor expand the log likelihood $\mathcal{L}$ to second order in the parameters about a fiducial point $\vec{\theta}_\mathrm{fid}$ to get
\begin{align}
\mathcal{L}(\vec{\theta};\vec{d}) &= \mathcal{L}(\vec{\theta}_\mathrm{fid};\vec{d}) + \delta\vec{\theta}^\intercal \nabla\mathcal{L}(\vec{\theta}_\mathrm{fid};\vec{d}) \label{eq:score} \\ \notag
&\hspace{1cm}+ \frac{1} {2}\delta\vec{\theta}^\intercal\nabla\nabla^\intercal\mathcal{L}(\vec{\theta}_\mathrm{fid};\vec{d})\delta\vec{\theta} +\mathcal{O}\big(\| \vec{\delta\theta\|}^3)   \ ,
\end{align}
where $\delta\vec{\theta}= \vec{\theta} -\vec{\theta_\mathrm{fid}}$ and $\| \cdot \|$ is the Frobenius norm, defined, for any matrix $\vec{x}$, as 
\begin{align}
 \|\vec{x}\|= \sqrt{\sum\limits_{i}\sum\limits_{j} |x_{ij}|^2}\ . 
\end{align} .

Taking just the linear term of Eq.~(\ref{eq:score}), the score 
provides a way of compressing the data to a vector whose length is equal to the number of parameters, and to linear order all information which the data has about the parameters is in the score.  If the likelihood is Gaussian and the covariance matrix does not depend on the parameters then score compression is equivalent to MOPED.

\cite{alsing2018generalized} showed that the score saturates the Cramér-Rao bound, so no other statistic provides more Fisher information. Thus if the likelihood is known, and a fiducial point can be chosen,  then compression to the score is optimal in the sense of retaining as much information as theoretically possible.  If the likelihood is not known exactly but can still be reliably approximated by a Taylor expansion around a  fiducial point, then score compression is again optimal, \textit{given the specific model}, but loses information compared to the situation where the true likelihood is known.\footnote{ \cite{alsing2018generalized} imply that the score is a sufficient statistic for the log likelihood but strictly this is only true locally in a Fisher-information sense.  The score is not necessarily sufficient globally. The same is true of MOPED compression.}  Of course if the likelihood cannot be estimated at all then score compression is not feasible.
\subsubsection{Linear and simple non-linear neural networks with Fisher-based loss function (LNN-FI and NN-FI)}\label{sect:LNN-FI}

We introduce here the idea of a simple linear neural network which we compare with MOPED.  The network has one input layer equal to the length of the uncompressed data vector and one output layer equal in length to the number of parameters. There are no non-linear features, in particular no activation functions.  Thus the compressed data $\vec{t}$ is just a weighted sum of the input data: \mbox{$\vec{t} = \vec{w}^\intercal \vec{d}$} for some weights $\vec{w}$.  The loss function is equal to the negative log determinant of the Fisher matrix with covariance and derivatives calculated at fixed fiducial parameter values. Thus the system has similarities to MOPED. However the weights are not necessarily the same as MOPED weights because the neural network is simply solving an optimisation problem, rather than finding an orthonormal set of basis functions.

 This loss function has the problem that the latent covariance, and hence the determinant of the Fisher matrix,  can increase indefinitely during training so that the loss never converges. To see this, suppose the weights found by the network are $\vec{w}$.  Then the compressed (latent) data is $\vec{t} = \vec{w}^\intercal\vec{d}$  and  the latent Fisher matrix is
\begin{align}
 \tens{F}_\mathrm{t} &= \frac{\partial{\vec{t}}}{\partial \theta}^\intercal\tens{C}_\mathrm{d}^{-1}\frac{\partial{\vec{t}}}{\partial \theta} \ ,
\end{align}
where $\tens{C}_\mathrm{d}$ is the covariance of the data.
But now if instead of $\vec{t} = \vec{w}^\intercal\vec{d}$ we have $\vec{t} = \alpha \, \vec{w}^\intercal\vec{d}$ for some constant $\alpha$, then the Fisher matrix scales as $\alpha^2$ and the loss has no maximum.

To avoid this problem we whiten the mean-subtracted data so that the input data covariance is equal to the identity. Thus 
\begin{align}
    \vec{d} \rightarrow \tilde{\vec{d} }=\tens{C}_\mathrm{d}^{-1/2}\vec{d}\ .
\end{align}
In this whitened space runaway covariance cannot occur. 
We show in Sect.~\ref{sect:FI_results} that such a network can in fact closely reproduce MOPED weights.

This idea can readily be expanded to include non-linear features  such as hidden layers and activation functions in the network, while retaining the same loss function.  We refer to such networks as NN-FI. The details of our implementation are given in Sect.~\ref{sect:FI_results}.

\subsubsection{Information-maximising neural network (IMNN)}\label{sect:IMNN}
IMNN  is a  non-linear method which maximises the Fisher information of the compressed data. Thus it can be considered to be an extension of score compression.  Unlike the other FI-based methods discussed here, IMNN makes no assumptions about the availability of derivatives or about the form or availability of the covariance matrix.  Instead these are learned from simulations as part of a neural network which finds a non-linear function $f_\phi$, with network parameters $\phi$, such that $\vec{t} = f_\phi(\vec{d})$ maximises the Fisher information.  To calculate the covariance IMNN requires many simulations at fixed fiducial cosmologies but with different initial conditions, and to calculate the derivatives it needs many simulations at different cosmologies. These are in addition to simulations needed to train the main network, making IMNN potentially very resource-intensive.

As with the linear network described in Sect.~\ref{sect:LNN-FI}, the covariance of the network output is scale invariant so the loss function needs careful construction to ensure that at every training epoch the covariance remains close to the identity.  Several approaches have been used to control the covariance, for example the originating (for cosmology) paper \cite{charnock2018automatic} has loss function:
\begin{align}
    \Lambda = -\mathrm{det}(\tens{F}) + \mathrm{det}(\tens{C}_f )\ ,
\end{align}
where $\tens{C}_f$ is the covariance of the network output. 
The extra term ensures that the covariance cannot stray too far from the identity.

In contrast the follow-up paper \cite{makinen2021lossless} has a more complex loss function
\begin{align}\label{eq:imnn-loss}
\Lambda = -\mathrm{det}(\tens{F}) + r_{\Lambda_C} \Lambda_C \ ,
\end{align}
where 
\begin{align}
    \Lambda_C = \frac{1}{2} \left( \|(\tens{C}_f-\mathbf{I})\|^2 + \|(\tens{C}^{-1}_f-\mathbf{I})\|^2 \right),
\end{align} 
and $r_{\Lambda_C}$ is a regularisation term given by
\begin{align}
    r_{\Lambda_C} = \frac{\lambda \Lambda_C}{\Lambda_C + \exp(-\alpha \Lambda_C)} \ ,
\end{align}
where $\lambda$ and $\alpha$ are user-defined parameters. 
Again, when the covariance is far from the identity the $r_{\Lambda_C}$ function helps bring it back.

As another example, \cite{prelogovic2024informative} used
\begin{align}  
    \Lambda &= -\ln\mathrm{det} (\tens{F}) \nonumber 
    +\lambda \, \|\tens{C}_f - \mathbf{I}\| \tanh (\|\tens{C}_f - \mathbf{I}\|)
    +\lambda_W \, L_2(W_{\mathrm{NN}}) \ .
\end{align}
  
The  second term regularises the covariance, with the $\tanh$ factor turning off regularisation as the covariance approaches unity.  $L_2(W_{\mathrm{NN}})$ is the Euclidean norm of the network weights, which regularises the weights.

From these examples, and comments in the papers, it can be seen that constructing a well-behaved loss function for IMNN can be difficult.

\subsection{Compression based on or related to mutual information}\label{sect:MI}
\subsubsection{Canonical correlation analysis (CCA)}\label{sect:CCA}

CCA is a standard technique  widely used in the social sciences but only recently introduced to cosmology by \cite{park2025dimensionality}.  The method finds linear combinations of two sets of random variables which are maximally correlated with each other. 
If the variables have Gaussian distributions, it also maximises the mutual information between the two.\footnote{Thus although we have categorised CCA as \lq MI-based\rq, this is not strictly true unless the two datasets have Gaussian distributions.} This is proved in Appendix \ref{sect:GaussianMI}.

The formalism of CCA is presented in many standard sources, but we explain it here in some detail for pedagogical reasons. In what follows we assume the general case where the distributions are not necessarily Gaussian and so mutual information is not necessarily optimised.

We want to create linear combinations of the elements of  vectors \mbox{$\vec{\theta} = (p_1,p_2,\ldots, p_n$)} and \mbox{$\vec{d} = (d_1,d_2,\ldots, d_m$)} which are maximally correlated. In general these could be any two vectors, but for data compression we are interested in the specific case where one vector is the vector of parameters and the other is the data.
So we want to find the weights $\vec{a}$ and $\vec{b}$ which maximise
\begin{equation}   
    \rho =\frac{\vec{a}^\intercal\tens{C}_{\mathrm{pd}}\vec{b}}{\sqrt{\vec{a}^{\intercal}\tens{C}_{\mathrm{p}} 
    \vec{a}}\sqrt{\vec{b}^\intercal\tens{C}_{\mathrm{d}} \vec{b}}} \ ,
\end{equation}
 where $\tens{C}_{\mathrm{p}}$ and $\tens{C}_{\mathrm{d}}$ are the covariances of the parameters and data respectively and $\tens{C}_{\mathrm{pd}}$ is their cross-covariance. Thus we maximise the correlation between the projected parameters $\vec{a}^\intercal\vec{\theta}$ and data $\vec{b}^\intercal\vec{d}$.

If we impose the constraints 
\begin{equation}\label{eq:constraints}
 \vec{a}^{\intercal}\tens{C}_{\mathrm{p}}\vec{a}=1  \ , \\
 \vec{b}^{\intercal}\tens{C}_{\mathrm{d}} \vec{b}=1 \ ,\\
\end{equation} 
then this turns into the problem of maximising $\vec{a}^{\intercal}\tens{C}_{\mathrm{pd}}\vec{b}$ subject to these constraints.  We define the Lagrangian, $\Lambda$, by
\begin{equation}
    \Lambda = \vec{a}^{\intercal}\tens{C}_{\mathrm{pd}}\vec{b} - \alpha (\vec{a}^{\intercal}\tens{C}_{\mathrm{p}}\vec{a}-1) - \beta (\vec{b}^{\intercal}\tens{C}_{\mathrm{d}} \vec{b}-1) \ ,
\end{equation}
where $\alpha$ and $\beta$ are Lagrangian multipliers. The derivatives of $\Lambda$ with respect to $\vec{a}$ and $\vec{b}$ are
\begin {align}
\frac{\partial\Lambda}{\partial\vec{a}}&= \tens{C}_{\mathrm{pd}}\vec{b} - 2\alpha \tens{C}_{\mathrm{p}}\vec{a}\\
\frac{\partial\Lambda}{\partial\vec{b}}&= \tens{C}_{\mathrm{dp}}\vec{a} - 2\beta \tens{C}_{\mathrm{d}}\vec{b} \ .
\end{align}
Setting the derivatives to zero gives
\begin{align}\label{eq:normal}
    \tens{C}_{\mathrm{pd}}\vec{b} &= 2\alpha \tens{C}_{\mathrm{p}}\vec{a}\\
    \tens{C}_{\mathrm{dp}}\vec{a} &= 2\beta\tens{C}_{\mathrm{d}}\vec{b} \ .
\end{align}
Then
\begin{align}
\vec{a}^\intercal\tens{C}_{\mathrm{pd}}\vec{b} &= 2\alpha \, \vec{a}^\intercal\tens{C}_{\mathrm{p}}\vec{a}= 2\alpha\\
    \vec{b}^\intercal\tens{C}_{\mathrm{dp}}\vec{a} &= 2\beta \, \vec{b}^\intercal\tens{C}_{\mathrm{d}}\vec{b}=2\beta \ ,
\end{align}
using the constraints in Eqs.~(\ref{eq:constraints}). It follows that $\alpha = \beta$, and we can choose to replace $2\alpha$ and $2\beta$ by $\gamma$. 

We then multiply Eqs.~(\ref{eq:normal}) by $\tens{C}_{\mathrm{p}}^{-1}$ and $\tens{C}_{\mathrm{d}}^{-1}$ respectively, giving
\begin {align}
 \tens{C}_{\mathrm{p}}^{-1}\tens{C}_{\mathrm{pd}}\vec{b} &= \gamma \,\vec{a}\label{eq:gammaa}\\
 \tens{C}_{\mathrm{d}}^{-1}\tens{C}_{\mathrm{dp}}\vec{a} &= \gamma \, \vec{b} \ .\label{eq:gammab}
\end{align}
We can substitute for $\vec{b}$ in Eq.~(\ref{eq:gammaa}), and for $\vec{a}$ in Eq.~(\ref{eq:gammab}) to get:
\begin{align} 
\tens{C}_{\mathrm{p}}^{-1}\tens{C}_{\mathrm{pd}}\tens{C}_{\mathrm{d}}^{-1}\tens{C}_{\mathrm{dp}}\vec{a} &= \gamma^2 \,\vec{a}\\ 
\tens{C}_{\mathrm{d}}^{-1}\tens{C}_{\mathrm{dp}}\tens{C}_{\mathrm{p}}^{-1}\tens{C}_{\mathrm{pd}}\vec{b} &= \gamma^2 \, \vec{b} \label{eq:gammasqb}\ . 
\end{align}
Thus $\vec{a}$ is an eigenvector of $\tens{C}_{\mathrm{p}}^{-1}\tens{C}_{\mathrm{pd}}\tens{C}_{\mathrm{d}}^{-1}\tens{C}_{\mathrm{dp}}$ and $\vec{b}$ is an eigenvector of $\tens{C}_{\mathrm{d}}^{-1}\tens{C}_{\mathrm{dp}}\tens{C}_{\mathrm{p}}^{-1}\tens{C}_{\mathrm{pd}}$, and these provide the optimal compression for our parameters and data respectively. In both cases the associated eigenvalue is $\lambda = \gamma^2$.

CCA is potentially an attractive compression method because it is a simple (linear) eigenvalue problem which  does not require derivatives. We discuss some practical issues with implementing CCA in Sect.~\ref{sect:MI_results}.

\subsubsection{Variational mutual information maximisation (VMIM)}\label{sect:VMIM}

VMIM was first used in cosmology by \cite{jeffrey2021likelihood}. By construction, it maximises the mutual information between the compressed data and the parameters, with the compression being achieved by a neural network $\vec{t}= f_\phi(\vec{d})$ with network parameters $\phi$. 

From Eq.~(\ref{eq:MIcond}), the definition of mutual information between the compressed data $\vec{t}$ and the parameters $\vec{\theta}$ is 
\begin{align}
    I(f_\phi(\vec{d});\vec{\theta})&=
 \mathbb{E}[\log P(\vec{\theta}|f_\phi(\vec{d}))] + H(\vec{\theta}) \ ,
\end{align}
where $P(\vec{\theta}|f_\phi(\vec{d}))$ is the conditional distribution of  the parameters and the compressed statistics.  In general this is unknown. However a lower bound to $P$ can be obtained with a neural density estimator trained to find a lower bound to $I(f_\phi(\vec{d});\vec{\theta})$.
In other words we find $q(\vec{\theta}|\vec{t};\varphi)$ with parameters $\varphi$ such that
\begin{align}
    I(f_\phi(\vec{d});\vec{\theta}) &\geq \mathbb{E}[\log q(\vec{\theta}|f_\phi(\vec{d});\varphi)] +H(\vec{\theta}) \\
    &\approx \frac{1}{N}\sum_i \log q(\theta_i|f_\phi(d_i);\varphi)+H(\vec{\theta})\ . 
\end{align}

$H(\vec{\theta})$ is independent of $\varphi$ and can be ignored so the problem becomes that of maximising $\log q(\vec{\theta}|f_\phi(\vec{d});\varphi)$ over $\phi$ and $\varphi$. This can be achieved by using neural density estimation to model $f_\phi(\vec{d})$ \citep{jeffrey2021likelihood, park2025dimensionality}. In a careful comparison of compression methods \cite{lanzieri2025optimal} showed empirically that it is possible for VMIM to retain all the information from the uncompressed data. However theoretically it only approaches losslessness if the lower bound to $P(\vec{\theta}|f_\phi(\vec{d}))$ is tight and the neural network architecture is flexible enough to produce the necessary lossless compression, conditions which are often difficult to meet in practice \citep{barber2004algorithm,poole2019variational}.

\subsubsection{Gaussian negative log likelihood (GNLL)}\label{sect:GNLL}

This method has a loss function of the general form
\begin{equation}
    \frac{1}{2} \log\big(\det (\tens{C_\mathrm{\theta}})\big) + \frac{1}{2}(\vec{t} - \vec{\theta})^{\intercal} \tens{C}_\mathrm{\theta}^{-1} (\vec{t} - \vec{\theta}) ,\label{eq:GNLLloss}
\end{equation}
where $\vec{t}$ is the compressed data (possibly the output of a neural network) and $\tens{C}_\mathrm{\theta}$ is the covariance matrix of the parameters $\vec{\theta}$. 
This was introduced to cosmology by \cite{perreault2017uncertainties} and \cite{fluri2018cosmological}, who explicitly included the parameter covariance to normalise the uncertainty scale.

GNLL can be considered as a special case of VMIM where instead of estimating $q(\vec{\theta}|f_\phi(\vec{d}))$, the analysis is simplified by taking $f_\phi(\vec{d})$ to be the output of a neural network and assuming that the posterior is Gaussian \citep{lanzieri2025optimal} so that \mbox{$q(\vec{\theta}) = \mathcal{N}\big(\vec{\theta};\vec{t}, \tens{C}_\theta\big)$}.

If the posterior is further restricted to have fixed variance then this method is equivalent to using a weighted  MSE loss function because 
\begin{align}
\log q(\vec{\theta})&= -\frac{1}{2}\sum_i(\theta_i-f_\phi(d_i))^\intercal 
\tens{C}_\theta^{-1}(\theta_i-f_\phi(d_i))\ .
\end{align}
This is demonstrated in appendix C of \citeauthor{lanzieri2025optimal} \citeyear{lanzieri2025optimal}.
 If the covariance is the identity the method becomes exactly equivalent to minimising the MSE of the predicted the parameters, which is discussed in Sect.~\ref{sect:NN-MSE}. 

As with the linear FI-based model described in Sect.~\ref{sect:LNN-FI}, we also consider a linear version of GNLL which we refer to as LNN-GNLL.  This uses the  loss function from Eq.~(\ref{eq:GNLLloss}) but the network has no hidden layers and no activation functions. 

\subsection{Compression based on mean squared error minimisation}\label{sect:other}
\subsubsection{Principal component analysis (PCA)}\label{sect:PCA}

PCA is a well-known method of reducing the size of a data vector $\vec{d}$ by  finding linear combinations which best explain the variance of the data. The methodology is standard and can be found in many sources. However we present it here to contrast it with other methods such as CCA which can also be tackled as eigenvalue problems.

We want to find a set of weights 
$\vec{b}$  such that \mbox{$\vec{t} =\vec{b}^\intercal\vec{d}$} projects the data onto a lower-dimensional space in a way which captures as much of the data variance as possible. 

The variance which we want to maximise is
$\vec{b}^\intercal\tens{C}_\mathrm{d}\vec{b}$. 
We can demand that the weight $\vec{b}$ is a unit vector so that \mbox{$\vec{b}^\intercal\vec{b}=1$}.  Maximising the variance subject to this constraint leads to the Lagrangian 
\begin{equation}
    \Lambda = \vec{b}^\intercal\tens{C}_\mathrm{d}\vec{b} - \lambda(\vec{b}^\intercal\vec{b}-1) \ .
\end{equation}
The derivative of $\Lambda$ with respect to $\vec{b}$ is
\begin{equation}
\frac{\partial\Lambda}{\partial\vec{b}} = 2 \tens{C}_\mathrm{d}\vec{b}  - 2\lambda  \vec{b} \ .
\end{equation}
Setting this equal to zero (for maximisation) gives
\begin{equation}
    \tens{C}_\mathrm{d}\vec{b}  = \lambda  \vec{b}\ .
\end{equation}
So $\vec{b}$ is an eigenvector of $\tens{C}_\mathrm{d}$ with eigenvalue $\lambda$, meaning that the linear combinations of the original data $\vec{d}$ which maximise the variance can be found from the eigenvectors of the covariance matrix. The eigenvalues quantify the proportion of variance explained by each eigenvector. If the eigenvalues are ordered so that $\lambda_1 > \lambda_2 > \lambda_3 \ldots$ then $\lambda_1$ explains the most variance, and so on. Often most of the variance is captured by only the first few eigenvectors, known as the principal components.

 We have categorised PCA as minimising MSE because maximising the variance is equivalent to minimising the mean squared reconstruction error, the difference between the uncompressed data and its reconstruction from the compressed data. 
To see this we use a property of the Frobenius norm to express the mean squared reconstruction error, $\|\vec{d} -\vec{b}\vec{b}^\intercal\vec{d}\|^2$, in terms of the trace of the error.
\begin{align}
    \|\vec{d} - \vec{b}\vec{b}^\intercal\vec{d}\|^2 &= \mathrm{Tr}[(\vec{d} - \vec{b}\vec{b}^\intercal\vec{d})^\intercal(\vec{d} - \vec{b}\vec{b}^\intercal\vec{d})] \\
    &= \mathrm{Tr}[(\vec{d}^\intercal - \vec{d}^\intercal\vec{b}\vec{b}^\intercal)(\vec{d}-\vec{b}\vec{b}^\intercal\vec{d})]  \\
    &= \mathrm{Tr}[\vec{d}^\intercal\vec{d} - 2(\vec{d}^\intercal\vec{b}\vec{b}^\intercal\vec{d})
    + \vec{d}^\intercal\vec{b}\vec{b}^\intercal\vec{b}\vec{b}^\intercal\vec{d}] \\
    &= \mathrm{Tr}[\vec{d}^\intercal\vec{d} - 2(\vec{b}^\intercal\vec{d})^2 +(\vec{b}^\intercal\vec{b})(\vec{b}^\intercal\vec{d})^2] \\
    &=\mathrm{Tr}[\vec{d}^\intercal\vec{d}] -\mathrm{Tr}[(\vec{b}^\intercal\vec{d})^2)] \label{eq:recon}\ ,   
\end{align}
where the last line uses the fact that $\vec{b}^\intercal\vec{b} = 1$.
  The first term of Eq.~(\ref{eq:recon}) is constant and the second is the trace of the variance of the compressed data.
  Thus to minimise the mean squared reconstruction error we need to maximise this variance, which is the optimisation criterion for PCA. 

PCA creates a more tractable dataset consisting of uncorrelated orthogonal linear combinations of the original data.  The new data vectors will hopefully describe or explain the data better than the original, and the size of the dataset can be reduced by discarding linear combinations which contribute little to the variance (those corresponding to small eigenvalues of the covariance).  Thus in a sense it is a compression method: it can be used to reduce the length of the data vector.
However, because PCA depends only on the data and not on the parameters there is no reason why the compressed data should be aligned with directions of greatest parameter sensitivity.  So in most cases PCA will lose both Fisher information and mutual information. 

 \cite{park2025dimensionality} suggested that the principal components can be made more sensitive to the parameters by using the average of the covariance at different points in parameter space rather than at a single fiducial point.  However this is only feasible if suitable covariances exist, and even then there is no guarantee that the compressed data will coincide with the  directions of maximum  sensitivity.  

\subsubsection{Neural network with mean square error loss of predicted parameters (NN-MSE)}\label{sect:NN-MSE}
A common approach to data compression is to minimise the mean squared error of the predicted parameters, on the grounds that this will retain those features of the data which best predict the parameters. Thus we train a neural network $f_\phi$ with parameters $\phi$ and loss function 
\begin{align}
    \frac{1}{n_\mathrm{\theta}}\sum_{i=1}^{n_\mathrm{\theta}} [f_\phi(d)_i - \theta_i]^2 \ ,
\end{align}
where $n_\mathrm{\theta}$ is the number of parameters and $f_\phi(d)_i$ is the network output.  This is equivalent to predicting the parameters from the data, or alternatively to estimating the mean of the posterior \citep{lanzieri2025optimal}.  The method can be implemented with one network per parameter, or with a single network which predicts all parameters.   

This method is simple to implement and has been used by, for example, \cite{jeffrey2021likelihood, jeffrey2025dark,novaes2024cosmological,novaes2025cosmology} as a prelude to simulation-based inference. However this loss function does not guarantee the retention of information: it is possible to predict the mean accurately without reproducing other features of the posterior distribution.  In Sect.~\ref{sect:MSE_results} we also consider the possibility of using a linear neural network with MSE loss (LNN-MSE) in the same spirit as  LNN-FI which was discussed in Sect.~{\ref{sect:LNN-FI}.
This is exactly equivalent to linear regression because the parameters are modelled as a linear function of the data and the loss function minimises the MSE  between the input and predicted parameters.

A related loss function uses the mean absolute error which can be shown to predict the median of the posterior.  It was used for example by \cite{ribli2019weak}. We do not consider this loss function further here.

\section{ Testing the performance of compression methods}\label{sect:tests}
To test the performance of these methods, we compare the cosmological constraints produced with MOPED, NN-FI, LNN-FI,  CCA, NN-GNLL, LNN-GNLL, PCA, NN-MSE and LNN-MSE.  We do not include VMIM in the comparison because it has the added complexity of requiring  a neural density estimation step.  This may be important when data compression is part of a simulation-based inference pipeline, but is less crucial when considering data compression as an end in itself, as we do here.  We also exclude more complicated neural networks with a Fisher information-based loss function since we found that for our data vector these did not improve on a linear network. We found IMNN to be particularly difficult to implement and unnecessarily complicated for our purposes.  

\subsection{Data}
In these tests our data consists of analytical weak lensing power spectra, derivatives and fiducial covariance.  Since we are only using the data for demonstration purposes, the setup is simple and not intended to correspond exactly to real survey data. The advantage of using such a basic example is that we do not need to worry about details of the data and can focus on the compression methods themselves. The disadvantage is that compression methods may behave differently for more realistic, complex data.  However for our pedagogical purposes simplicity is preferable.

We consider four cosmological parameters, $\Omega_\mathrm{m}$, $\sigma_8$, $w_0$ and $w_a$, and assume a flat Universe.
We assume a \textit{Euclid}-like survey with area 15\,000~deg$^2$, total galaxy density 30~arcmin$^{-2}$ and redshift range $0.0 \le z \le 2.0$. The overall redshift probability distribution of source galaxies is 
\begin{align}
p(z)&\propto z^\alpha \exp\left[-\left(\frac{z}{z_0}\right)^\beta\right]\ ,\label{eq:pz}
\end{align}
with $\alpha = 2.0$, $\beta = 1.5$, $z_0 =z_\mathrm{med}/ \!\sqrt{\mathstrut 2}$, $z_\mathrm{med}=0.8$.
We use only three redshift bins, each containing the same number of galaxies. This results in redshift bin boundaries  $[0.02,0.81]$, $[0.81,1.18]$ and $[1.18,2.00]$. 

We model statistical uncertainty in photometric redshift values by assuming that the redshift distribution within each tomographic bin is Gaussian with a dispersion $\sigma_\mathrm{ph}$. Thus the conditional probability of obtaining a photometric redshift $z_\mathrm{ph}$ given the true redshift $z$ has the form
\begin{align}
p(z_\mathrm{ph}|z)\propto \exp\left[-\frac{\left(z_\mathrm{ph}-z\right)^2}{2\sigma^2_\mathrm{ph}\left(1+z\right)^2}\right]\ ,
\end{align}
where we take  $\sigma_\mathrm{ph}$ to be 0.05. 
We use 20 angular bins equally logarithmically spaced from \mbox{$\ell_\mathrm{min} =30$} to \mbox{$\ell_\mathrm{max} =3000$}. This produces a power spectrum data vector with 120 entries.

The tomographic weak lensing power spectrum is calculated in the standard way  as
\begin{align}
 	C^{ij}(\ell) 	&= \int_0^{\chi_\mathrm{lim}}\mathrm{d}\chi \ q^{(i)}(\chi)\,q^{(j)}(\chi)\, \chi^{-2} P_\delta\left(k; \chi\right)\ ,\label{eq:WLPS}
\end{align}
where $P_\delta$ is the non-linear matter power spectrum,  the weight $q^{(i)}(\chi)$ is given by
\begin{align}
	q^{(i)}(\chi)&= \frac{3H_0^2\Omega_\mathrm{m}}{2c^2}\frac{\chi}{a(\chi)} \int _\chi^{\chi_{\mathrm{lim}}}\mathrm{d}\chi^\prime \ p^{(i)}(\chi^\prime)\frac{\left(\chi^\prime-\chi\right)}{\chi^\prime} \ ,\label{eq:q}
\end{align}
and we make the Limber and flat-sky approximations. 
The power spectrum covariance is consistent with the methodology in \cite{reischke2025kids}. We use only the Gaussian term, given by
\begin{align}
	\mathrm{Cov}[\hat{C}^{ij}(\ell_1),\hat{C}^{mn}(\ell_2)]_\mathrm{G}
	&=\frac{\delta^K_{\ell_1\ell_2}}{2\ell_1\Delta\ell_1f_\mathrm{sky}}
	[C^{im}(\ell_1)C^{jn}(\ell_2)  \notag\\
    & \hspace{1.6cm} +C^{in}(\ell_1)C^{jm}(\ell_2)] ,
\end{align}
where  $\delta^K_{\ell_1\ell_2}$ is the Kronecker delta, and \mbox{$\langle\hat{C}^{ij}(\ell)\rangle \equiv{C}^{ij}(\ell) $}.  
This results in a block-diagonal covariance matrix in $\ell$ with off-diagonal entries within each $\ell-$block.
We calculate the derivatives of the power spectrum with respect to the parameters using finite differences with a five-point stencil.
\begin{table}
    \centering
     \caption{Priors used in the main illustrative analysis}
    \label{tab:priors}
    \begin{tabular}{c l S[table-format=-1.1]}
    \toprule
    \textbf{Parameter} &
    \multicolumn{1}{c}{\textbf{Prior}} &
    {\textbf{Fiducial value}}\\
    \midrule
    $\Omega_\mathrm{m}$ & $\mathcal{U}[0.1,0.4]$ & 0.32\\
    $\sigma_8$          & $\mathcal{U}[0.6,1.0]$ & 0.81\\
    $w_0$               & $\mathcal{U}[-2.0,0.4]$ & -1.00\\
    $w_a$               & $\mathcal{U}[-1.0,1.0]$ & 0.00\\
    \bottomrule
    \end{tabular}
   
\end{table}

We generate 5000 weak lensing power spectrum data vectors spanning the priors shown in Table~\ref{tab:priors}, together with  associated analytical derivatives and covariance at the fiducial point which is also shown in Table~\ref{tab:priors}.  Samples are selected using a Latin hypercube over the prior ranges. For some  analyses we use variants of these selection criteria, or smaller samples; we point out below when this is the case.  We add Gaussian noise generated from the fiducial covariance to each data vector.  For the neural network models, we divide the data into samples of 4000 for training and 1000 for validation.

\subsection{Neural network architecture and hyperparameters}
The neural networks which we use are all essentially autoencoders with different loss functions.  We train for up to 1000 epochs in each case, with early stopping to prevent over-fitting. We keep most hyperparameters fixed. In particular we use the Adam optimiser and  batch size of 64. We experimented with different learning rates between 0.0001 and 0.05 and as a result use a rate of 0.001 throughout.

All networks have an input layer equal to the length of the datavector and an output layer with size  equal to the number of cosmological parameters.  Our simple linear networks have no activation functions and no hidden layers. For each multi-layer network, after experimentation,  we use four hidden layers of lengths \mbox{[64,32,16,8]}. The input layer and first three hidden layers are followed by Leaky ReLU activation functions.  We found changing this architecture made little difference to results and was suitable for all the models we considered.

\subsection{Markov chain Monte Carlo (MCMC) analysis }

After compressing the data with each of the methods, we use the cosmological parameter estimation code \texttt{CosmoSIS}\footnote{https://cosmosis.readthedocs.io/en/latest/} \citep{zuntz2015cosmosis} to perform MCMC sampling using the package \texttt{emcee} \citep{foreman2013emcee}, assuming in each case that the likelihood is sufficiently Gaussian.

\subsection{Figure of merit}\label{sect:FoM}
We use a single figure of merit (FoM) to compare  compression methods: the log determinant of the Fisher matrix of the compressed data, a proxy for the log of the inverse of the posterior volume.  To calculate this we use the same fiducial covariance and derivatives throughout, including for methods not based on an FI loss function.   This FoM measures how well the compression retains information about all parameters jointly, close to the fiducial point.  We checked that it does not penalise methods which do not explicitly maximise Fisher information by also calculating the parameter covariance from the MCMC chains. For all methods except PCA (where the posterior volume from the MCMC chains is likely prior-limited), the negative log determinant of the covariance from the chains was close to the value obtained using the Fisher matrix. 
We also explored using 2D FoMs in the $\Omega_\mathrm{m}-\sigma_8$ and $w_0-w_a$ planes but found that these had too much randomness to allow useful comparisons between compression methods.

\section{Results}\label{sect:results}
\subsection{FI-based loss results}\label{sect:FI_results}

The FI-based loss functions which we compare are principally MOPED and LNN-FI, the linear network described in Sect.~\ref{sect:LNN-FI} which has no hidden layers or activation functions. We also briefly consider NN-FI.
 
 As a check on the performance of LNN-FI we first initialised the network  with the exact weights derived from MOPED and confirmed that it could closely reproduce the MOPED weights.  
If more realistically we initialise the network with random weights then the output weights are still close to MOPED as shown in Fig.~\ref{fig:weights2}. This shows  the fractional contributions of the absolute values of the weights for the parameter $\Omega_\mathrm{m}$, i.e.
   ${|w_i|}/{\sum_i |w_i|} $, averaged over ten sets of randomly-initialised weights.  We obtained similar agreement for the other three parameters.

\begin{figure}
    \centering
    \includegraphics[width=8cm]{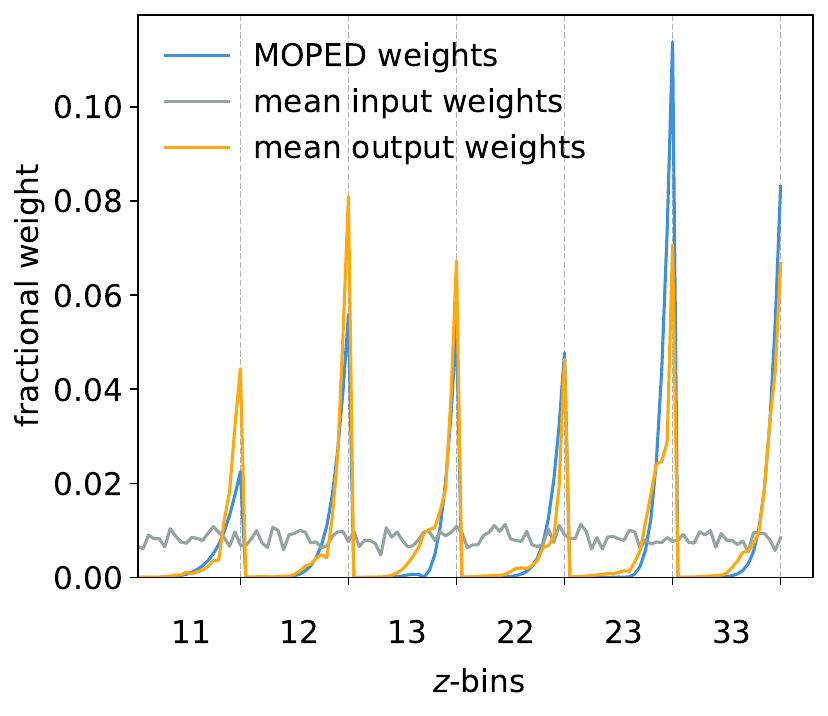}
    \caption{Comparison between linear network output weights  and MOPED weights for $\Omega_\mathrm{m}$, taking the average of the output weights over 10 iterations of random initialisation. \label{fig:weights2}}  
\end{figure}

Figure~\ref{fig:contoursFI} shows parameter constraints from these  compression methods. 
The linear and non-linear neural networks produce virtually identical constraints to the uncompressed data, as expected for our Gaussian data. The FoM for MOPED was identical to that for uncompressed data (as is analytically guaranteed to be true) and the two neural networks achieved similar FoMs.

\begin{figure}
\centering
 \includegraphics[width=9cm]{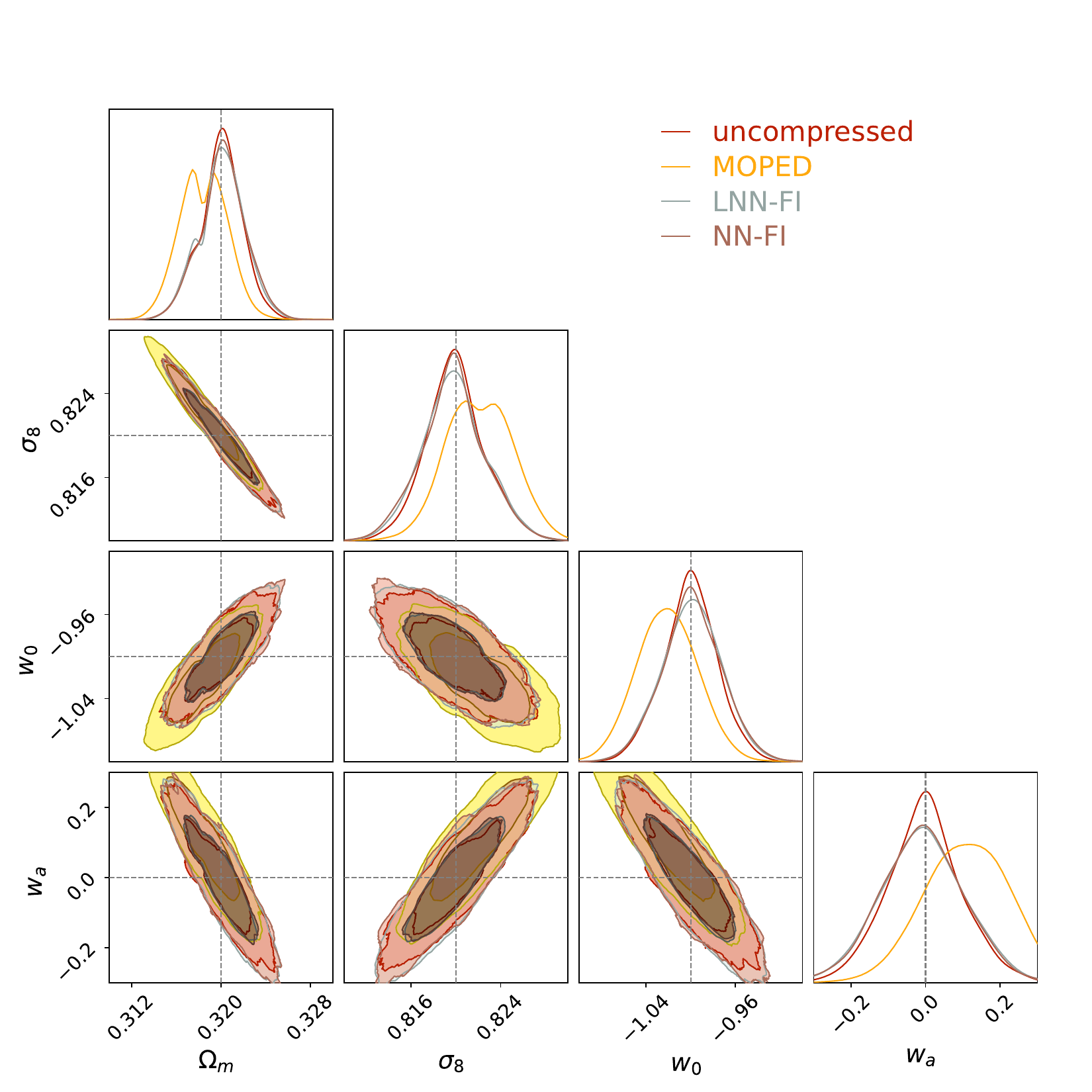}
    \caption{ Parameter constraints obtained with compression methods based on FI. The constraints obtained with uncompressed data are shown as a benchmark. True (fiducial) values are marked by grey lines. \label{fig:contoursFI}} 
\end{figure}

We investigated whether a more complex neural network, NN-FI, can improve on MOPED, by introducing non-linearity in a controlled way, first by adding a single hidden layer of size 64 followed by a Leaky ReLU activation function, and then by  varying the size of this layer. However we found that this non-linear network did not extract any additional information (see Fig.~\ref{fig:contoursFI}) and so we did not pursue the possibility of a more complex architecture in this case.

\subsubsection{How \lq  correct\rq \ does the fiducial point need to be for MOPED?}\label{sect:MOPEDfid}
We found in Sect.~\ref{sect:FI_results} that both MOPED and a simple linear neural network can compress our data  losslessly with FoM equal to that of uncompressed data.  However
in both cases,  and indeed with any Fisher information- based neural network, we require a covariance matrix and derivatives at a fiducial point in parameter space and it may not be known \textit{a priori} what that point should be. \cite{heavens2000massive} showed that this may not be a serious problem in practice, because it is always possible to iterate to identify a suitable fiducial point. Similarly \cite{lin2023simulation} showed that score compression can be robust to a non-optimal choice of fiducial cosmology.  However, this still leaves  questions about how much Fisher information may be lost if the fiducial point is \lq  wrong\rq, what causes this information loss, and therefore when it may become a problem.

To answer these questions we consider how much information is lost if we use the MOPED weights $ w(\theta_\mathrm{fid})$, calculated at the supposed fiducial point, to produce  \lq transported\rq \ Fisher matrices $\tens{F}_\mathrm{trans}(\theta_i) $ at a set of different points $\theta_i$.  We then compare the determinant of $\tens{F}_\mathrm{trans}(\theta_i) $ with that of  $\tens{F}_\mathrm{full}(\theta_i)$ calculated with the uncompressed data at $\theta_i$.  (For simplicity we use the same fiducial covariance in both cases.) This leads to the metric
\begin{align}
    \Delta (\theta_i) &= \log \mathrm{det} [\tens{F}_\mathrm{trans}(\theta_i;w_\mathrm{fid}) ]- \log \mathrm{det}[\tens{F}_\mathrm{full}(\theta_i;w_\mathrm{full}) ] \ .
\end{align}
A value of $\Delta (\theta_i)$ close to zero suggests that little information is lost.  However this is not sufficient, which can be understood intuitively by considering two 2-dimensional Fisher ellipses corresponding to the same determinant value. The orientation and eccentricity of the ellipses also matter.  
We therefore need to delve further into the causes of information loss. In particular we want to discover whether information loss is due to parameter degeneracies or because the  tangent plane at $\theta_i$ is  oriented differently from that at $\theta_\mathrm{fid}$, and which parameters are affected most. 

To probe the reasons for information loss we consider the generalised eigenvalue problem \citep{ghojogh2019eigenvalue}
\begin{align}
 \tens{F}_\mathrm{trans} \vec{v} &= \lambda \ \tens{F}_\mathrm{full}\vec{v}  
\end{align}
or equivalently
\begin{align}
 {\tens{F}^{-1}_\mathrm{full}}\tens{F}^{ }_\mathrm{trans} \vec{v} &= \lambda \vec{v}  \ .
\end{align}

We now define the relative Fisher information matrix between  $\tens{F}_\mathrm{trans}$ and
$\tens{F}_\mathrm{full}$ as
\begin{align}
\tens{R}&\coloneq
{\tens{F}}^{-1/2}_{\mathrm{full}}
{\tens{F}}^{ }_{\mathrm{trans}}
{\tens{F}}^{-1/2}_{\mathrm{full}}   \ .
\end{align}

This symmetric matrix 
quantifies the relative information content of $\tens{F}_\mathrm{trans}$ compared to $\tens{F}_\mathrm{full}$. This is a concept often used in the field of optimal design of experiments to predict how much more informative one experiment will be than another \citep{pukelsheim2006optimal}.

The eigenvalues of $\tens{R}$ identify whether $\tens{F}_\mathrm{trans}$ retains all the information of $\tens{F}_\mathrm{full}$ in each of the principal  directions (linear combinations of parameters) when the MOPED weights are transported from $\theta_\mathrm{fid}$ to $\theta_i$.  If an eigenvalue is close to unity then most information is retained in the relevant direction.  An eigenvalue less than one indicates loss of information in that direction.  It is also useful to examine the eigenvectors associated with eigenvalues which are less than one. These eigenvectors represent linear combinations of parameters which lose information. A parameter with  a large weight in the eigenvector is badly affected by use of inappropriate weights.

Another possible diagnostic is the principal angles between the tangent planes at $\theta_\mathrm{fid}$ and $\theta_i$ \citep{bjorck1973numerical, golub2013}. These are the angles  in each direction in data space required to rotate one tangent plane onto the other.  They vary from zero to $\SI{90}{\degree}$ and measure how tilted the derivatives at $\theta_i$ are compared to those at $\theta_\mathrm{fid}$. Thus they indicate whether loss of information is at least partly due to the geometry of the setup. 

It can be shown \citep{bjorck1973numerical} that for orthogonal projections (as with MOPED) and whitened data, the eigenvalues of $\tens{R}$ are related to the principal angles $\alpha_j$ by
\begin{align}
    \lambda_j = \cos^2 (\alpha_j) \ . \label{eq:prinangle}
\end{align}
In practice the relationship given by Eq.~(\ref{eq:prinangle}) may be almost linear over the range of principal angles considered and these angles may not offer much additional insight. However it may be intuitively easier to think  about rotation angles rather than eigenvalues.

Table~\ref{tab:diagnostics} summarises these diagnostics and what can be learned from them. 

It can be useful to plot these diagnostics against a measure of the distance separating each point from the fiducial point.  This distance can be measured in either data space or parameter space; which to use is largely a subjective choice. 

In data space we use the metric
\begin{align}
    s^2_\mathrm{data} &= (\vec{d}-\vec{d}_\mathrm{fid})^\intercal\tens{C}^{-1}_\mathrm{fid}(\vec{d}-\vec{d}_\mathrm{fid}) \ .
\end{align}
In whitened data space where the covariance is equal to the identity this becomes
\begin{align}
     s^2_\mathrm{data} &= (\vec{d}-\vec{d}_\mathrm{fid})^\intercal(\vec{d}-\vec{d}_\mathrm{fid}) \ .
\end{align}
In parameter space we use 
\begin{align}
     s^2_\mathrm{par} &= (\vec{\theta}-\vec{\theta}_\mathrm{fid})^\intercal\tens{F}_\mathrm{fid}(\vec{\theta}-\vec{\theta}_\mathrm{fid})\ .
\end{align}

\begin{table*}
    \centering
    \setlength{\tabcolsep}{8pt}
     \caption{Diagnostics for assessing information loss when an inappropriate fiducial point is chosen }
    \label{tab:diagnostics}
    \begin{tabular}{ll}
    \toprule
   \multicolumn{1}{c}{\textbf{Diagnostic}} &
    \multicolumn{1}{c}{\textbf{Purpose}} \\
    
    \midrule
    
        relative information matrix $\tens{R} = \tens{F}_\mathrm{full}^{-1/2}\tens{F}^{  }_\mathrm{trans}\tens{F}_\mathrm{full}^{-1/2}$ & relative information content of $\tens{F}_\mathrm{full}$ and $\tens{F}_\mathrm{trans}$\\[0.5em] 
          eigenvalues of $\tens{R}$, $\lambda_j$ &if $\lambda_j \approx 1$, little information loss \\
           & if $\lambda_j \ll 1$, information loss \\[0.5em] 
          difference in log det of Fisher matrix at 2 points \\\hspace{2.5cm}vs distance between points&whether distance is the main problem  \\[0.5em] 
          eigenvalues vs distance metric &whether geometry is the main problem\\[0.5em] 
          principal angles  vs distance metric &whether geometry is the main problem\\[0.5em] 
  
         relative weights of parameters in eigenvector & which parameters are affected by information loss in direction of eigenvector \\[0.5em] 
        shape and orientation of Fisher ellipses& effect on parameter constraints\\\\
    \bottomrule     
    \end{tabular}   

\end{table*} 

\subsubsection{Example based on weak lensing power spectrum}
To illustrate the diagnostic tools described in Sect.~\ref{sect:MOPEDfid}, we use our weak lensing data to explore the information loss when the \lq  wrong\rq \ derivatives and hence \lq  wrong\rq \ MOPED weights are used at a non-fiducial point. This is purely to  show how the techniques can be used; the results are particular to this data and would not apply in other cases.

We choose a sample of 200 points $\theta_i$ relatively close to the maximum likelihood point of our posterior obtained from MCMC, selected using a Latin hypercube from the $5\sigma$ region of the posterior. This results in the (very tight) ranges: $\Omega_\mathrm{m} [0.315, 0.325]$,  $\sigma_8 [0.812,0.826]$, $ w_0 [-1.06,-0.93]$, $w_a [-0.301,0.230]$.  We use the same fiducial point as in Table~\ref{tab:priors}.

We first examine the eigenvalues \mbox{$\lambda_j, j = 0,\ldots,3$} of the matrix $\tens{R}$ for each point in the sample. In Fig.~\ref{fig:evMchi2} we plot these against the distances of the points from the fiducial point in data space. All except $\lambda_0$ are consistently close to one, indicating that  information is retained. In contrast $\lambda_0$ is often much less than one, showing that information is lost in the direction of the corresponding eigenvector.  However there is no clear relationship between the value of $\lambda_0$ and the distance from the fiducial point.  This is explored further in  Figs.~\ref{fig:deltalogdetchsq} and \ref{fig:deltalogdetdsq} which plot the difference between the log determinants of the true and transported Fisher matrices against the distances in data space and parameter space respectively.  In these plots the points are colour-coded by the value of $\lambda_0$, showing that the difference in the log determinants is related to the value of $\lambda_0$, rather than to the distance from the fiducial point.  This suggests that the loss of information may be due to geometry.  To confirm this, in Fig.~\ref{fig:prin_angles} we plot the difference in the log determinants of the Fisher matrices against the corresponding principal angles associated with each eigenvalue. For $\lambda_0$ there is a clear relationship between  information loss and principal angle, indicating that the loss of information is due to geometric factors.  There is no such relationship for the other eigenvalues which always have relatively small principal angles.

We can investigate  which parameters are most affected by this loss of information by examining their weighting in the eigenvector which corresponds to $\lambda_0$.  Figure~\ref{fig:evec_weights} shows the weighting given to each parameter in this eigenvector at each of our sample points. This clearly shows that the most problematic parameter in this case is $w_a$.  In our specific case this is not unexpected, since we already know that $w_a$ is not well constrained by the data.  However for the other three parameters little information is lost if the \lq wrong\rq \ fiducial point is used for MOPED.

\begin{figure}
    \centering
    \includegraphics[width=8cm]{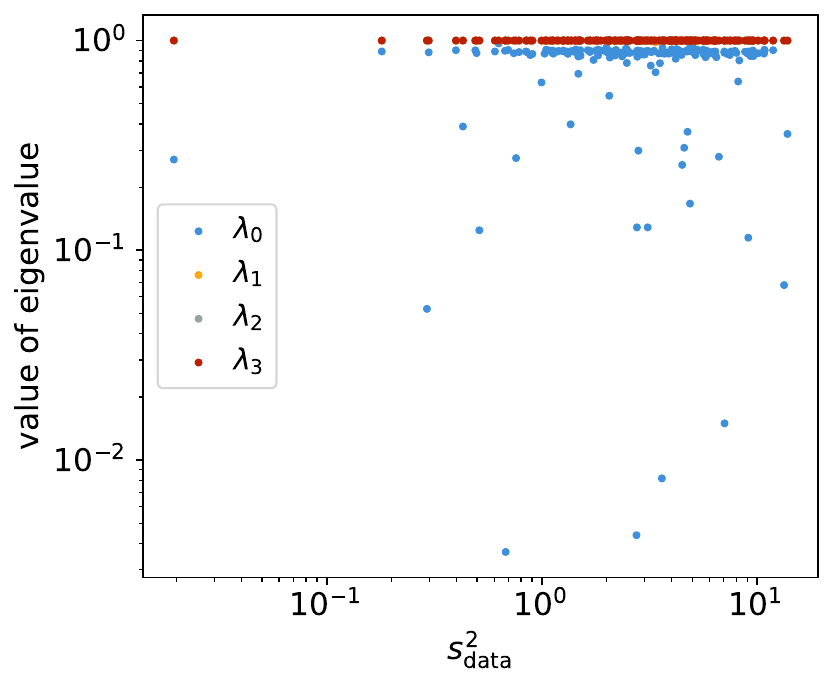}
    \caption{Values of the eigenvalues of the matrix $\tens{R}$ at at a sample of 200 points, plotted against the squared distance $s^2_\mathrm{data}$ from the fiducial point in data space. Eigenvalues are distinguished by colour. All eigenvalues except $\lambda_0$ are very close to 1 and cannot be distinguished by eye.\label{fig:evMchi2}} 
\end{figure}

\begin{figure}
    \centering
    \includegraphics[width=9cm]{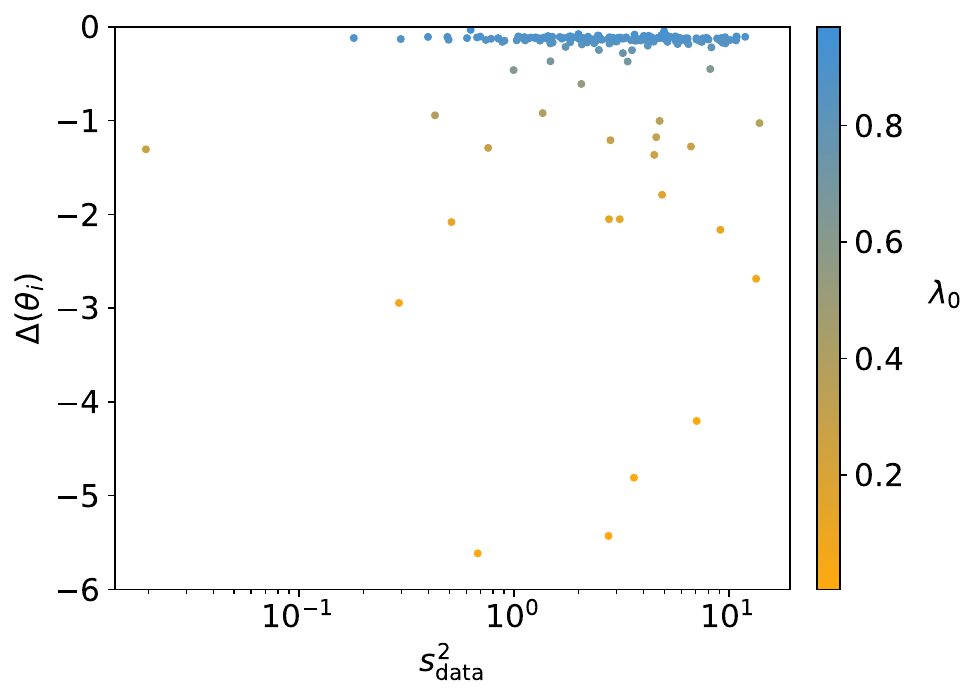}
    \caption{The difference $\Delta(\theta_i)$ between the log determinant of the Fisher matrix calculated with weights and derivatives at the fiducial point and with \lq  true\rq \ weights and derivatives at a sample of 200 points, plotted against the  distance from the fiducial point in data space. Colours indicate the value of the smallest eigenvalue, $\lambda_0$, of the matrix $\tens{R}$.\label{fig:deltalogdetchsq}} 
\end{figure}

\begin{figure}
    \includegraphics[width=9cm]{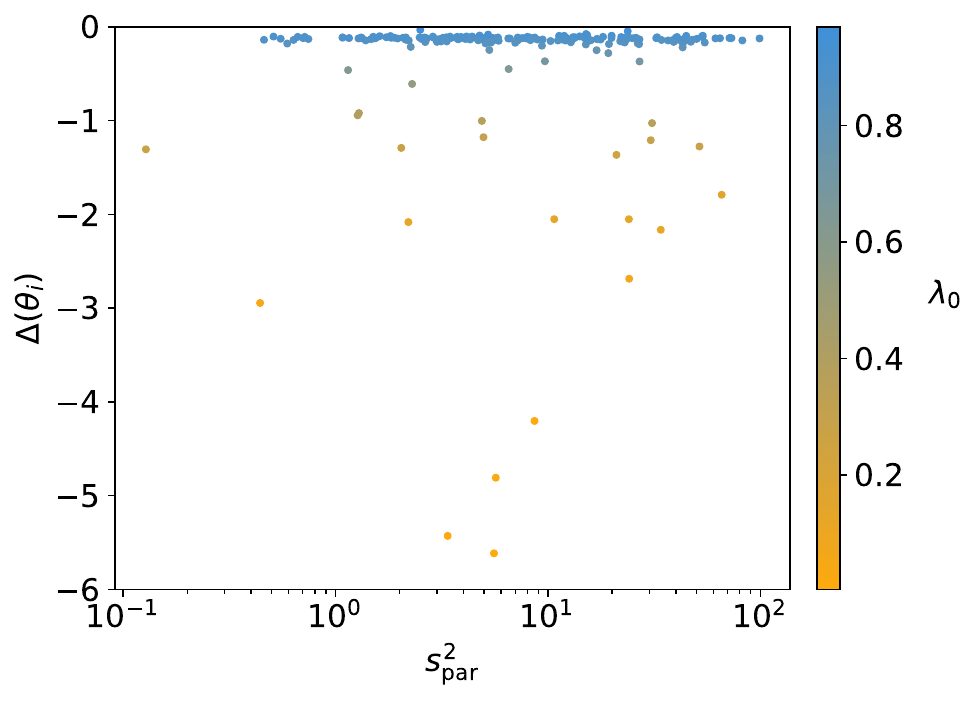}
    \caption{ The difference $\Delta(\theta_i)$ between the log determinant of the Fisher matrix calculated with weights and derivatives at the fiducial point and with \lq  true\rq \ weights and derivatives at a sample of 200 points, plotted against the distance from the fiducial point in parameter space. Colours indicate the value of the smallest eigenvalue, $\lambda_0$, of the matrix $\tens{R}$.  \label{fig:deltalogdetdsq}}
\end{figure}

\begin{figure}
    \centering
    \includegraphics[width=9cm]{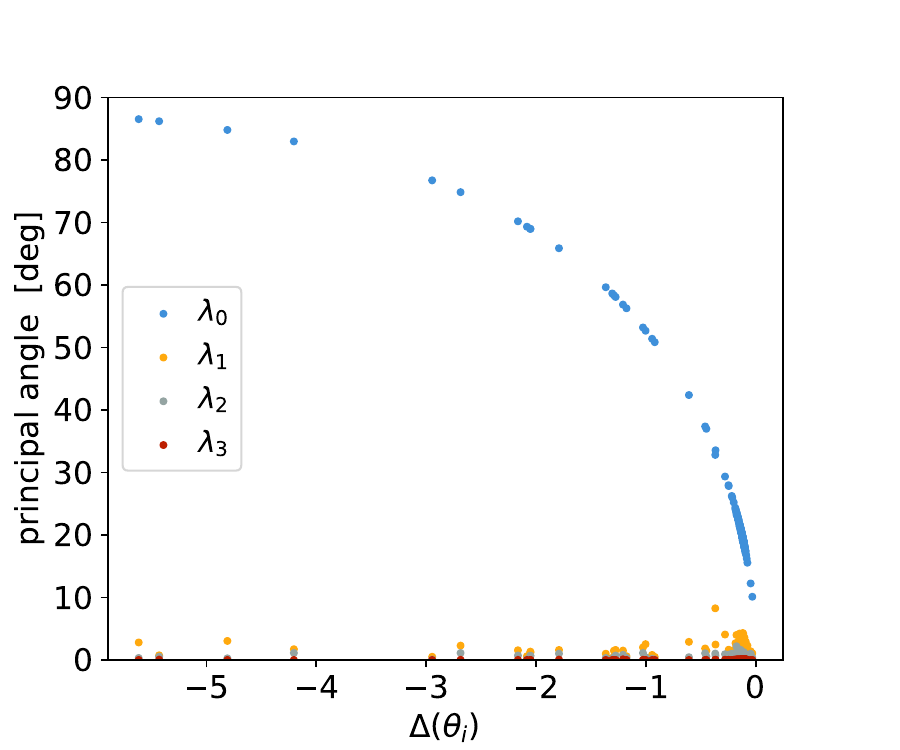}
    \caption{Principal angles plotted against the difference $\Delta(\theta_i)$ between the log determinant of the Fisher matrix calculated with weights and derivatives at the fiducial point and with \lq  true\rq \ weights and derivatives at  a sample of 200 points in parameter space. Eigenvalues are distinguished by colour.}\label{fig:prin_angles} 
\end{figure}
\begin{figure}
    \centering
    \includegraphics[width=8cm]{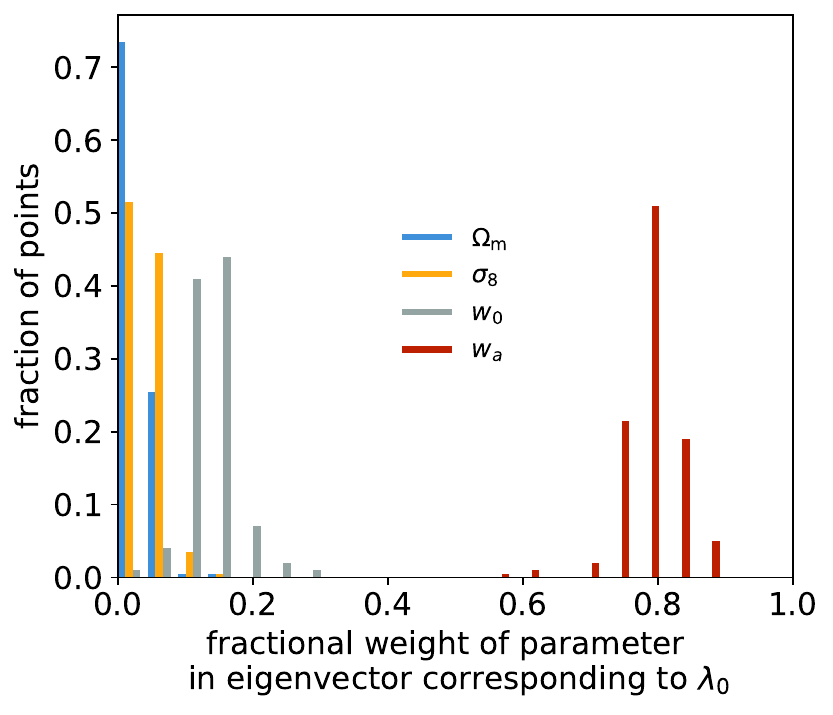}
    \caption{Weighting of parameters in the eigenvector associated with the smallest eigenvalue of $\tens{R}$, $\lambda_0$, for a sample of 200 points from the $5\sigma$ region of the posterior. }\label{fig:evec_weights} 
\end{figure}

Finally it can also be instructive to examine Fisher ellipses for \lq good\rq \ and \lq bad\rq \ points as illustrated by Fig.~\ref{fig:FMellipses}. This shows the  ellipses in the $w_0-w_a$ plane obtained for two arbitrarily chosen points, one where transport of weights does not lose information, and one where it does.
\begin{figure*}
  \centering
  
    \includegraphics[width=14cm]{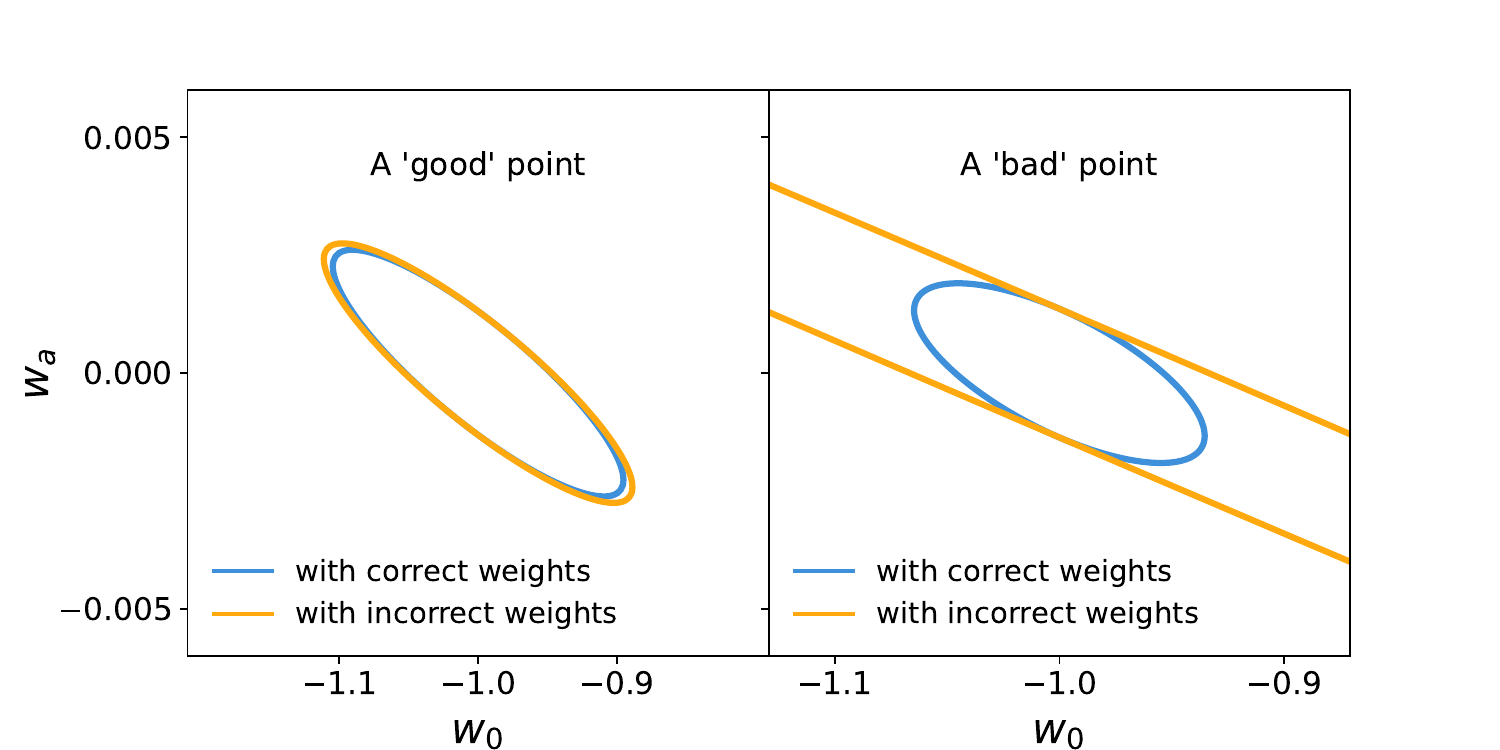}
  \caption{Fisher ellipses in the $w_0-w_a$ plane using  correct weights evaluated at an appropriate point and incorrect weights evaluated at the \lq wrong\rq \ point for two illustrative points in parameter space.  \textit{Left}:  a \lq  good\rq \ point where information is not lost by using the  incorrect compression weights. \textit{Right}: a \lq  bad\rq \ point where using the incorrect weights loses information. }
  \label{fig:FMellipses}
\end{figure*}

\subsection{MI-based loss results}\label{sect:MI_results}

\begin{figure}
 \includegraphics[width=9cm]{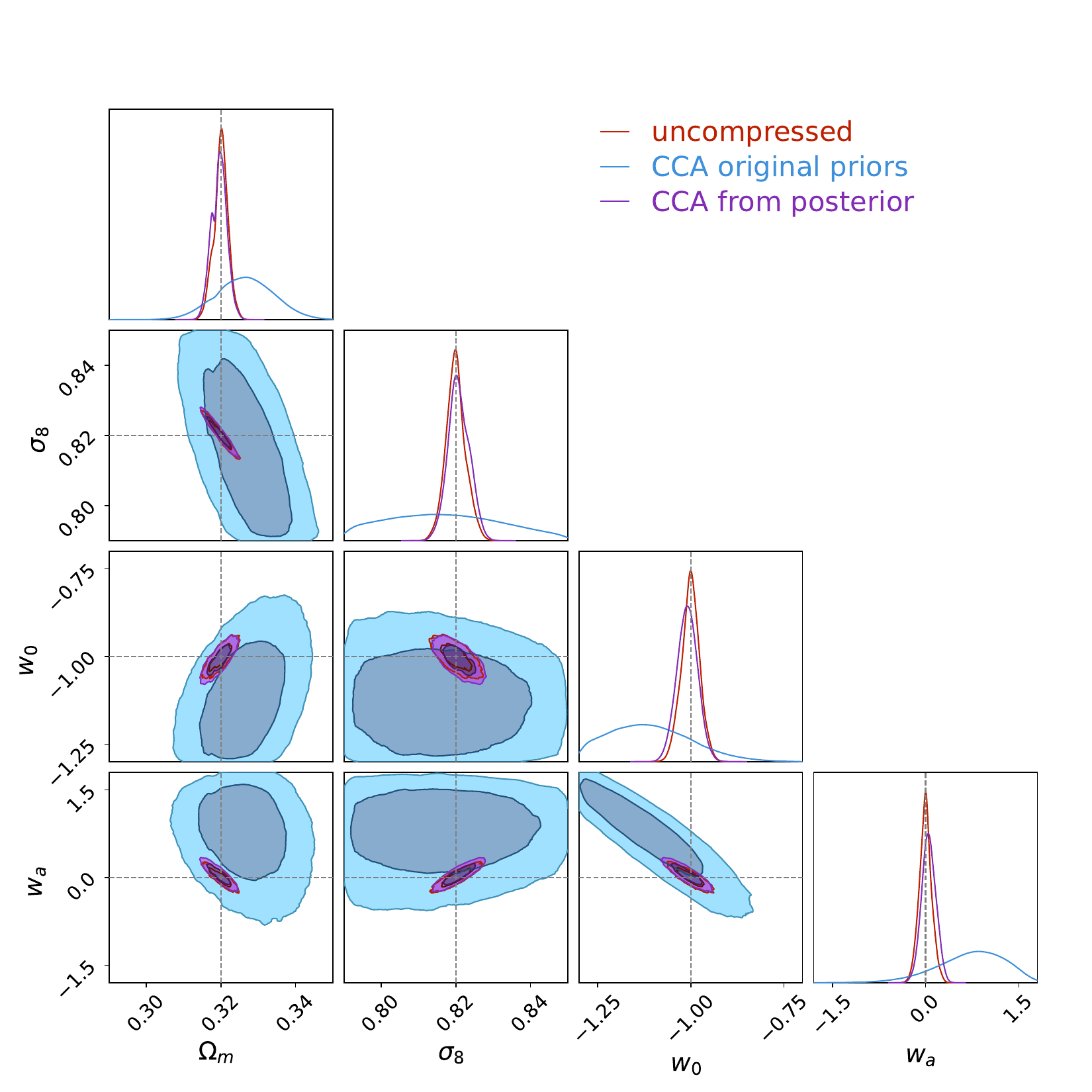}
    \caption{ Parameter constraints obtained with CCA, using either wide priors from Table~\ref{tab:priors} or narrow priors from the $1\sigma$ region of the original posterior.  The constraints obtained with uncompressed data are shown as a benchmark. True (fiducial) values are marked by grey lines.}\label{fig:contoursCCA} 
\end{figure}

We first report our results for CCA because, as Fig.~\ref{fig:contoursCCA} shows, when we used the wide priors set out in Table~\ref{tab:priors} we obtained very poor  parameter constraints (the blue contours in the figure).  
 However we found that if instead we select the sample points from a prior based on the posterior from our MCMC analysis of the uncompressed data, then we achieve much tighter parameter constraints, approaching those obtained with uncompressed data. This is shown  by the purple contours in Fig.~\ref{fig:contoursCCA} where the samples are selected from a prior based on the $1\sigma$ region of the original posterior (which was obtained with wide priors). We also experimented with using a prior region selected from the $0.5\sigma$ region of the posterior but this did not significantly change the results.  Finally we tried an iterative process starting from the original posterior, using this posterior to create the priors for the second iteration, then using the second posterior for the priors of a third iteration, and so on. However we found that because the first iteration produced constraints close to those based on uncompressed data, little was gained in the later iterations.
 
 Thus we conclude that a good strategy is to obtain a new set of priors by first running MCMC with a broad but reasonable choice of priors, making the (possibly incorrect) assumption that the likelihood is Gaussian, and if necessary using only a rough estimate of the covariance.  From this we  obtain an estimate of the posterior, and then run CCA with priors selected from this estimate, with the $1\sigma$ region of the posterior being a reasonable choice of prior.   This strategy is similar in spirit to the two-stage process used by \cite{park2025dimensionality} for their CCA compression.  They derived the likelihood by neural density estimation, then sampled from this for their MCMC analysis, and showed that this produced constraints similar to those from MOPED.

\begin{figure}
    \centering
    \includegraphics[width=9cm]{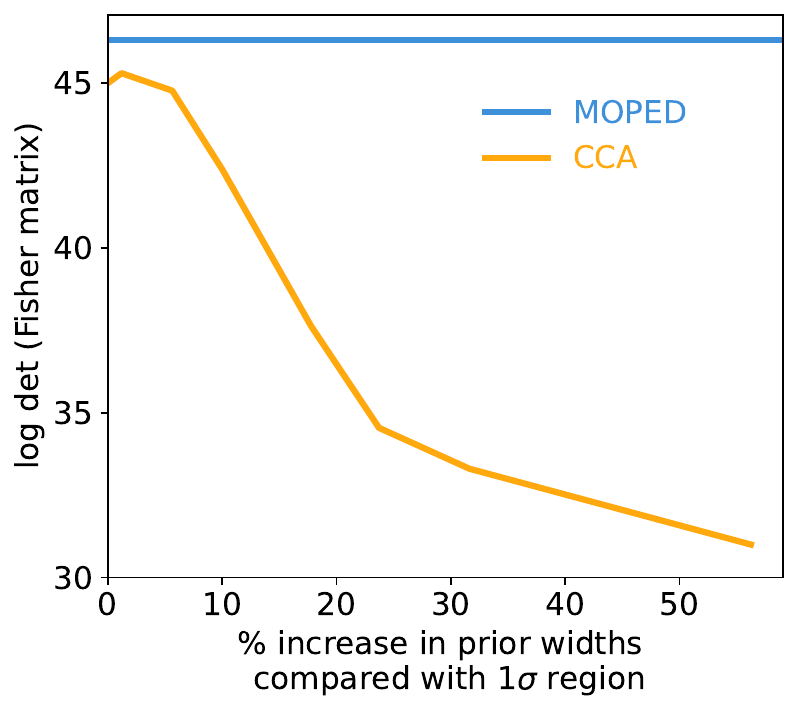}
    \caption{The log determinant of the Fisher matrix plotted against size of prior for CCA. The $x$-axis shows the percentage increase in the linear dimensions of the region from which samples are selected, with zero equal to the $1\sigma$ region of the posterior. The value of the log determinant achieved with MOPED is shown for comparison.}\label{fig:CCA_priors} 
\end{figure}
\begin{figure}
    \centering
    \includegraphics[width=9cm]{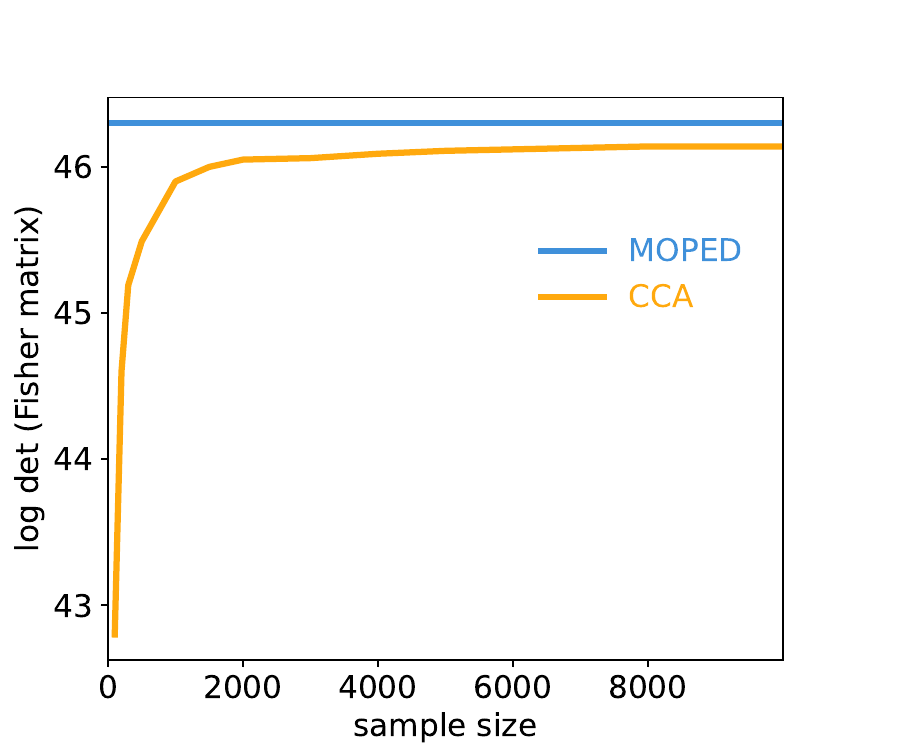}
    \caption{Relationship between log determinant of the Fisher matrix for CCA and the number of data points sampled. The value of the log determinant achieved with MOPED is shown for comparison. The vertical scale is different from  Fig.~\ref{fig:CCA_priors}. }\label{fig:CCA_samples} 
\end{figure}

To emphasise how the constraints from CCA depend on the priors, Fig.~\ref{fig:CCA_priors} shows how the log determinant of the Fisher matrix changes as the priors are widened.  The $x$-axis shows the percentage increase in the linear size of the prior ranges, with zero corresponding to the $1\sigma$ region of the posterior. At each step all parameter ranges are increased by the same percentage.  If the prior region is small then information retention is close to MOPED, but subsequently falls steeply before beginning to level off when the increase in priors is around 30--50 per cent. Our original  priors in Table~\ref{tab:priors} are even wider than used for this figure which explains the poor compression which we originally obtained with CCA. 

A larger sample of points is also useful. Figure~\ref{fig:CCA_samples}, in which points are sampled from the $1\sigma$ region of the posterior, shows that information retention increases rapidly as the sample size grows but then levels off, suggesting an optimum sample size. (The vertical scale is different from  Fig.~\ref{fig:CCA_priors}). 

Having shown that CCA can produce compression which is close to MOPED, in Fig.~\ref{fig:contoursMItight} we compare CCA (sampling from the initial posterior) with other MI-based compression methods.  The linear neural network with no hidden layers, LNN-GNLL, performs poorly against our FoM and in this case we added non-linearity and increased the number of hidden layers in the stepwise manner described in Sect.~\ref{sect:FI_results} to create our GNLL model.  We  obtained good results from a network with four hidden layers with sizes \mbox{[64, 32, 16,8] } with Leaky ReLU activation functions after the first three layers and a learning rate of 0.001.  Adding more hidden layers or changing the layer sizes did not improve on this.

We also applied the same two-stage method as for CCA to these methods and again obtained improvements, in particular for LNN-GNLL, although unlike with CCA the results still did not approach the MOPED FoM.
\begin{figure}
 \includegraphics[width=9cm]{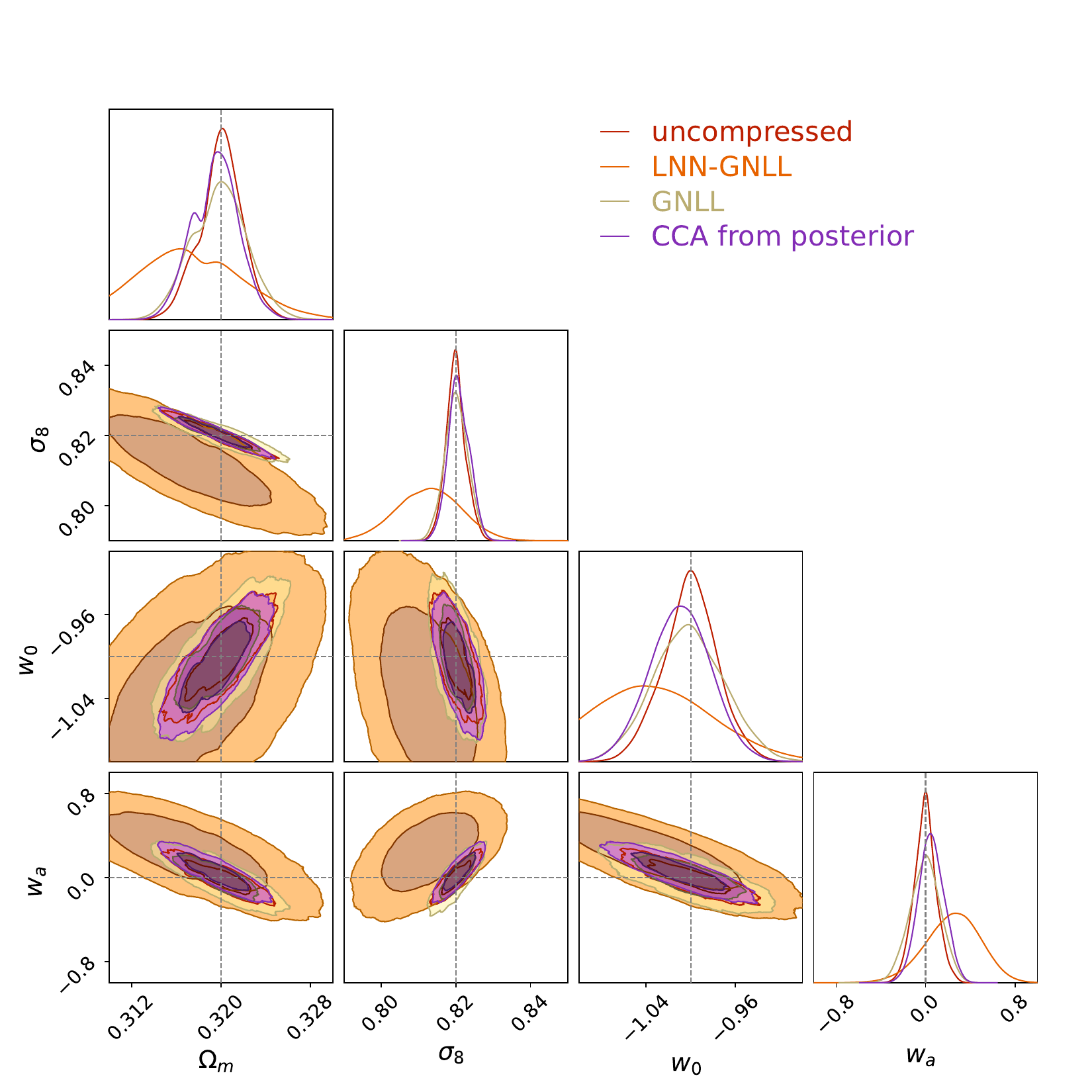}
    \caption{ Parameter constraints obtained with compression methods based on MI. The constraints obtained with uncompressed data are shown as a benchmark. True (fiducial) values are marked by grey lines. \label{fig:contoursMItight}} 
\end{figure}

\begin{figure}
    \centering
    \includegraphics[width=9cm]{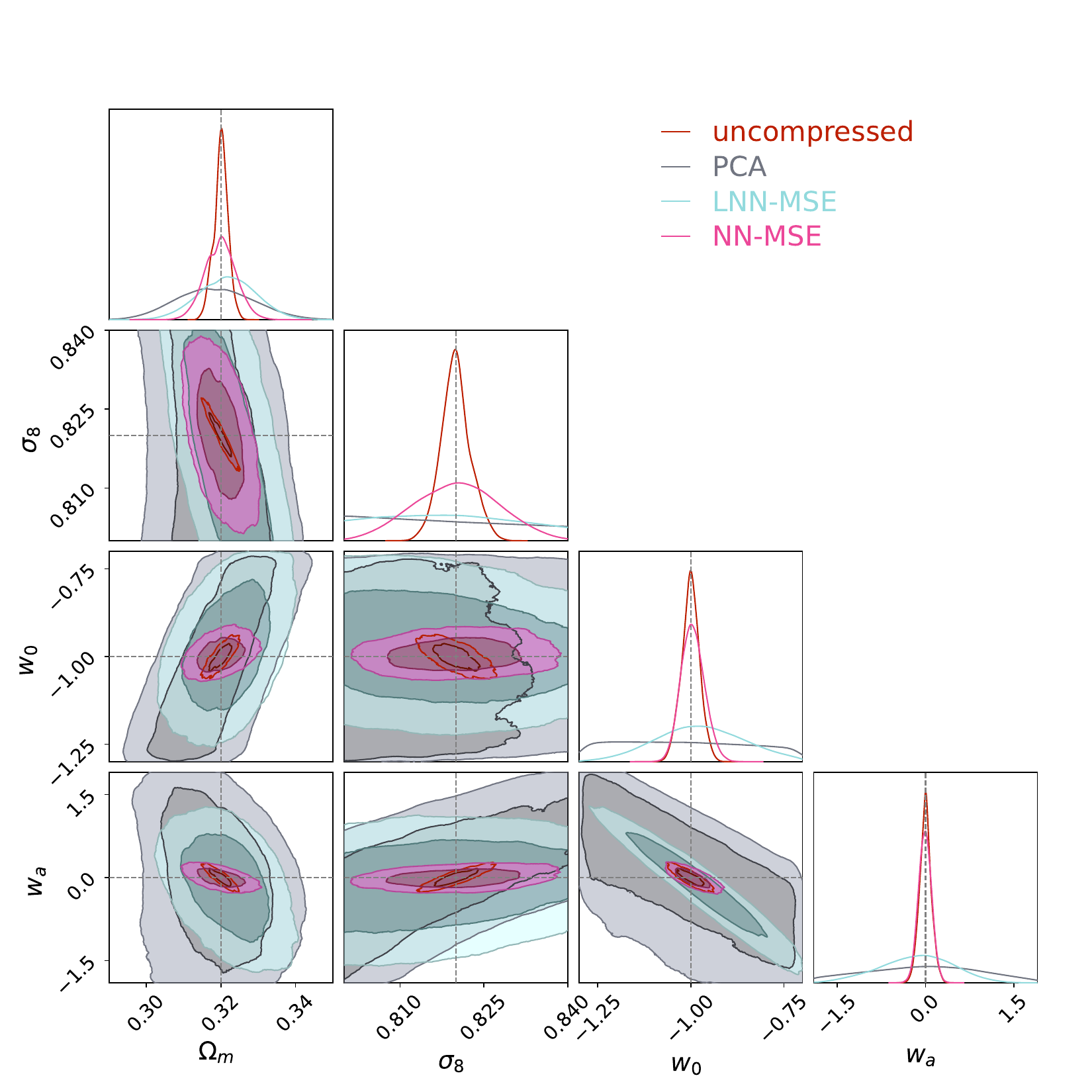}
    \caption{Parameter constraints obtained with compression methods which are based on MSE-based loss functions. The constraints obtained with uncompressed data are shown as a benchmark. True (fiducial) values are marked by grey lines.\label{fig:contoursother}}  
\end{figure}
\subsection{Results from MSE-based loss functions}\label{sect:MSE_results}
Figure~\ref{fig:contoursother} shows constraints obtained when MSE loss criteria are used.  In all cases information is lost, and the orientation of the contours is not consistent with that from the uncompressed data. 

PCA performs particularly poorly. This could be because for Fig.~\ref{fig:contoursother} we took only four principal components, for comparison with other methods which compress to the number of cosmological parameters. We tested whether retaining more principal components would help, but in fact most of the variance is captured by only two eigenvalues.  We also experimented with the two-stage process which we used for CCA but this did not improve the FoM. 

The results for neural networks with MSE loss were obtained with single networks which predict all parameters. We found that using a separate network for each parameter, as was done, for example, by \cite{park2025dimensionality} and \cite{novaes2025cosmology}, did not significantly improve the prediction performance of either LNN-MSE or NN-MSE.  As with NN-GNNL, described in Sect.~\ref{sect:MI_results}, we experimented with increasing the number and sizes of layers in the NN-MSE network but could not improve on the same architecture that we used for NN-GNNL, that is four hidden layers with sizes [64, 32, 16, 8] with Leaky ReLU activation functions after the first four layers and a learning rate of 0.001.  (We used the same structure in both cases for simplicity; there is no theoretical reason why both should be the same).   

Once again we also applied the two-stage method described in Sect.~\ref{sect:MI_results} to methods with MSE loss functions and again obtained improved FoMs, especially for the linear neural network. Even so, these methods still fell well short of MOPED.

\begin{figure*}
\centering
 \includegraphics[width=12cm]{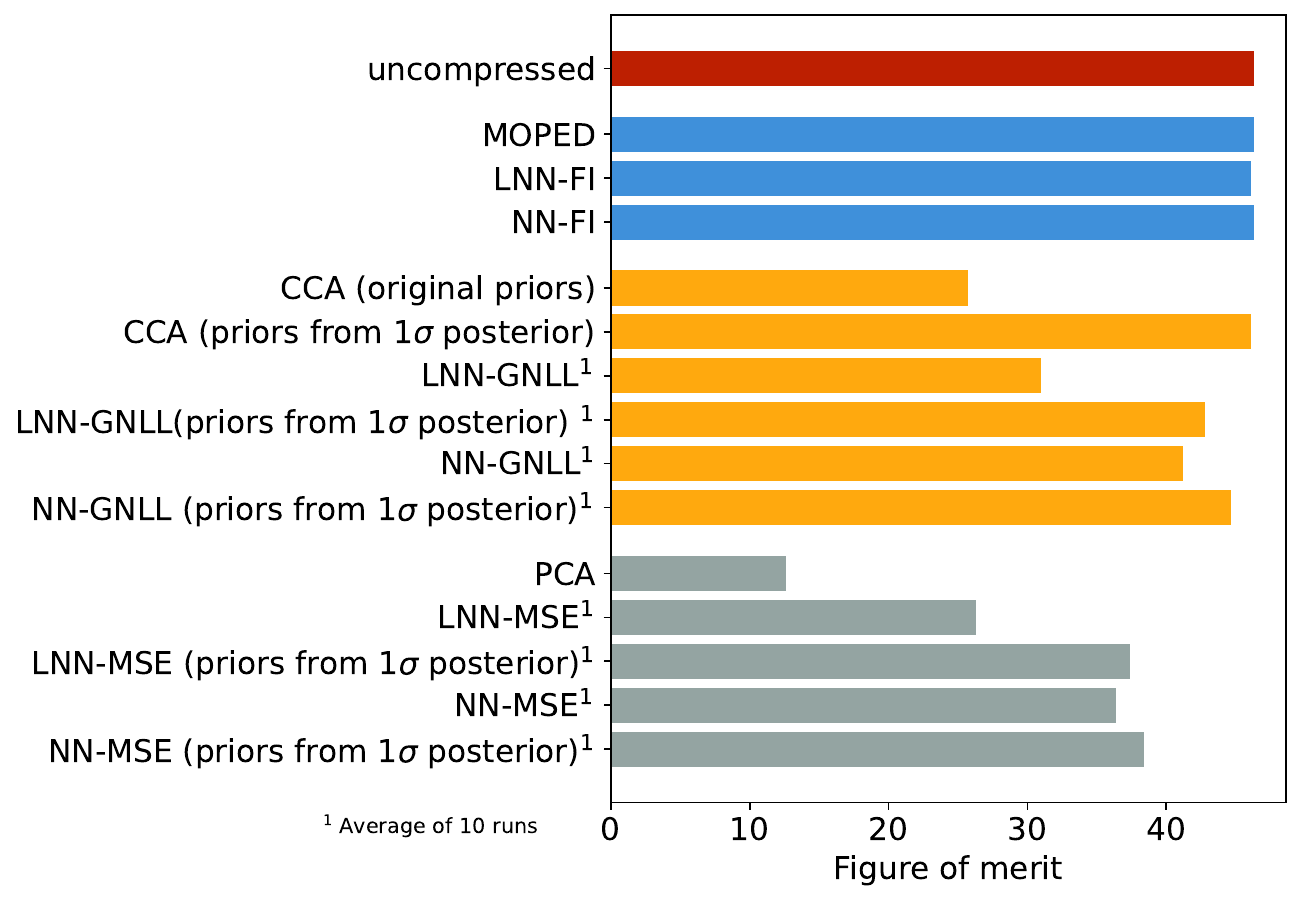}
    \caption{ The values of the figure of merit, the log determinant of the Fisher matrix, obtained with each compression method and with uncompressed data. Methods labelled \lq priors from $1\sigma$ posterior\rq \ are obtained using the two-stage process described in Sect.~\ref{sect:MI_results}. High-level descriptions of the  compression methods can be found in Table~\ref{tab:compression_methods}.}
\label{fig:summary_results}
\end{figure*}

\subsection{Summary of results}\label{sect:summary_results}
Figure~\ref{fig:summary_results} shows the values of our figure of merit, the log determinant of the Fisher matrix, for each method and for uncompressed data. As discussed in Sect.~\ref{sect:FoM}, this FoM approximates the covariance of the full posterior well.  This figure shows that: 
\begin{itemize}
    \item all methods with Fisher information-based loss functions were close to lossless;
    \item CCA can be improved by sampling from the $1\sigma$ region of an approximate posterior, achieving almost the same FoM as MOPED;
    \item a non-linear network with mutual information-based loss function performed no better than a similar linear network;
    \item if the loss function is based on minimising the mean squared error of the parameters, a non-linear network is required to achieve the same information retention as methods based on mutual information;
     \item the same two-stage method used for CCA also improves the performance of other methods with MI- and MSE-based loss functions, but they are still not close to the MOPED result.

\end{itemize}

\section{Discussion and conclusions}\label{sect:conclusions}
Data compression is often essential for simulation-based inference and for numerical estimation of large covariance matrices, and may also be important in other situations where data vectors are large.  In Sect.~\ref{sect:methods} we described several methods for compressing data, including those which have most commonly been used for cosmological analyses.  We divided the methods into those which maximise the Fisher information (FI) of the compressed data, those which maximise mutual information (in which we include canonical correlation analysis even though it only maximises mutual information under restricted conditions), and those which do not have an information-based loss criterion but are based on minimising mean squared errors. We used the weak lensing power spectrum as a straightforward, well-understood example data vector to explore the performance of most of these methods, omitting those which proved to be unnecessarily complex. We also elucidated connections between the methods; these are not always intuitively obvious.

We were particularly interested in investigating the need for methods based on complex neural networks,  and also whether it is possible to get good compression in cases where the true values of the parameters are not known with any certainty so it is not possible to define a suitable fiducial cosmology (necessary for FI-based methods).

We confirmed that for our relatively well-behaved Gaussian setup, compression based on Fisher information performs well if we know how to choose the fiducial cosmology.  This is expected as numerous studies have shown that MOPED or compression to the score work well.  We also showed that a linear neural network with FI loss can effectively reproduce MOPED, and that with our data nothing was gained from using a non-linear network.

Recognising that it may be desirable to use MOPED or score compression even if the fiducial cosmology is not known well, we described diagnostics which can be used to explore how far away from the fiducial point MOPED weights based on fiducial values can still yield useful results. This involved examining the eigenvalues and eigenvectors of the relative information matrix $\tens{R}$ which quantifies how much  Fisher information is retained when compression weights are transported from the fiducial point, relative to the Fisher information of the uncompressed data (or equivalently, to MOPED at the fiducial point). We demonstrated the method with our power spectrum data and drew conclusions about which cosmological parameters suffered most. In our particular case $w_a$ was most affected by a non-optimal choice of fiducial cosmology. However, although the diagnostic method is general, the results will depend on the specific dataset used. 

We showed that compression methods related to mutual information, which consider the whole posterior volume, can provide tight parameter constraints, but may require a two-stage process. In particular canonical correlation analysis (CCA) did not perform well if the data was drawn from wide priors (corresponding to lack of knowledge of where the bulk of the posterior mass is concentrated). However CCA produced better results when it was run twice, once to obtain a broad estimate of the posterior from  MCMC analysis, and then using samples drawn from the high-probability part of this posterior. Importantly, the original run can be based on a quite imprecise estimate of the covariance, can assume the likelihood is Gaussian, and needs no certain knowledge of where the posterior mass is concentrated. Using this strategy we found that the performance of CCA was close to that of MOPED if samples were drawn from a region ten per cent of the volume of the $1\sigma$ region of the original posterior, but fell away quickly if the priors were wider.  We also found that CCA performance increased with the size of the sample, up to a value at which it plateaued. 
The same two-stage strategy also improved the FoM of other MI-based methods, but unlike CCA these still did not approach the MOPED FoM.

Principal components analysis (PCA) and methods which minimise the MSE of the predicted parameters performed relatively poorly, even when we applied the two-stage strategy.  PCA is not generally a useful compression method for parameter inference since it does not take account of the parameters at all. It only produces tight  constraints if the principal components happen to be in the directions of maximum parameter sensitivity. Of course it is still a valuable method for understanding and simplifying complex data.  Other linear and non-linear methods with an MSE-based loss function are easy to use but were not lossless with our data.  We do not recommend them because  in many circumstances MOPED, CCA, or linear FI- based neural networks are just as easy to implement and retain more information.

We found the more sophisticated non-linear methods Variational mutual information maximisation and Information-maximising neural networks difficult to implement.  Since simple linear or non-linear methods worked well with our data and require fewer simulations, we saw no advantage in complicating the analysis. Of course it does not follow that non-linear methods are unnecessary for more complex situations such as field-level inference, very non-Gaussian data or analyses involving many nuisance parameters. As we noted in Sect.~\ref{sect:VMIM}, \cite{lanzieri2025optimal} found that VMIM was able to retain almost all information in a field-level analysis, and this can perhaps be considered the gold-standard method for complex data vectors, even if it is unnecessary for a two-point analysis.

Our overall recommendation is not to unnecessarily over-complicate. Linear methods such as MOPED, CCA and linear neural networks can provide good compression in the right circumstances, and it is possible to overcome the problem of not knowing \textit{a priori} where the posterior is concentrated.

\section*{Acknowledgements}

We thank Marco Gatti for helpful advice.
The authors acknowledge support by the ERC-selected UKRI Frontier Research Grant EP/Y03015X/1.
We acknowledge use of Chain Consumer \citep{Hinton2016} to produce contour plots and \cite{petroff2021accessible} for guidance on accessible plot colours. We used OpenAI’s ChatGPT (GPT-5.5) to explore concepts, to generate small self-contained pieces of code, and to proof-read equations.

\section*{Conflicts of interest}
The authors declare no conflict of interest.
\section*{Data Availability}
The data underlying this article will be shared on reasonable request to the corresponding author.
\bibliographystyle{rasti}
\bibliography{refs}

\appendix
\section{Mutual information, Kullback--Leibler divergence and Fisher information}\label{sect:MI_FI}
A connection can be seen between mutual information and Fisher information through the Kullback–-Leibler (KL) divergence. The Fisher information matrix is the variance of the score, and it can be shown that the score has mean zero, so
\begin{align}
    \tens{\tens{F}}_{ij} &= -\mathbb{E}\bigg [\frac{\partial\mathcal{L}(\vec{x}|\vec{\theta})}{\partial \theta_i}\frac{\partial\mathcal{L}(\vec{x}|\vec{\theta})}{\partial \theta_j}\bigg] \ ,
\end{align}
where $\mathcal{L}(\vec{x}|\vec{\theta}) $ is the log likelihood: $\mathcal{L}(\vec{x}|\vec{\theta}) = \log L(\vec{x}|\vec{\theta})$ for likelihood $L(\vec{x}|\vec{\theta})$.

If $\mathcal{L}$ is well-behaved,  e.g. it is twice differentiable, then 
\begin{align}
    \tens{\tens{F}}(\vec{\theta}) &= \mathbb{E}\bigg [\frac{\partial^2 \mathcal{L}(\vec{x}|\vec{\theta})}{\partial \vec{\theta}^2 }\bigg] \\
    &= -\int \mathrm{d}\vec{x} \ L(\vec{x}|\vec{\theta)}\Bigg[\frac{\partial\log L(\vec{x}|\vec{\theta)}}{\partial\vec{\theta}}\Bigg]^2 \label{eq:Fisher}\ .
\end{align}

A relationship between the Fisher information and KL divergence can be derived by considering the KL divergence between the likelihoods $L(\vec{x}|\vec{\theta}) $ and $L(\vec{x}|\vec{\theta+\delta\theta}) $  evaluated at different parameter values \citep{cover1991information,lehoucq2022entropy}.
From the definition of the KL divergence in Eq.~(\ref{eq:KLdiv}),
\begin{align}
    \mathrm{KL}(L(\vec{x}|\vec{\theta})\|L(\vec{x}|\vec{\theta+\delta\theta} ))
    =\int   \mathrm{d}\vec{x} \ L(\vec{x}|\vec{\theta})\log \frac{L(\vec{x}|\vec{\theta})}{L(\vec{x}|\vec{\theta+\delta\theta)}} \ ,\label{eq:KLdefn}
\end{align}
where the integration is only over $\vec{x}$ since $\vec{\theta}$ is a parameter of the distribution rather than a random variable.

For brevity define \mbox{$K(\vec{\delta\theta}) = \mathrm{KL}(L(\vec{x}|\vec{\theta})\|L(\vec{x}|\vec{(\theta+\delta\theta}) $}) and Taylor-expand this around $\delta\vec{\theta}=0$ to get
\begin{align}
  K({\delta\theta})
  &= K(0) + \vec{\delta\theta}^\intercal\frac{\partial K(\vec{\theta},\vec{\theta+\delta\theta)} }{\partial \vec{\theta}}\Bigg|_{\vec{\delta\theta}=0} \label{eq:KLpq}\\ \notag
 & \hspace{2cm}+ \frac{1}{2}\vec{\delta\theta}^\intercal\frac{\partial^2 K(\vec{\theta},\vec{\theta+\delta\theta) }}{\partial \vec{\theta}^2}\vec{\delta\theta}\Bigg|_{\vec{\delta\theta}=0} 
 + \mathcal{O}\big(\| \vec{\delta\theta\|}^3) \\
  &=\frac{1}{2}\vec{\delta\theta}^\intercal\frac{\partial^2 K(\vec{\theta},\vec{\theta+\delta\theta) }}{\partial \vec{\theta}^2}\vec{\delta\theta}\Bigg|_{\vec{\delta\theta}=0}
 + \mathcal{O}\big(\| \vec{\delta\theta}\|^3)
\label{eq:KLapprox} \ .
\end{align}
The last line follows because the first two terms of Eq.~ (\ref{eq:KLpq}) are zero: the first by definition and the second because $K(\vec{\theta},\vec{\theta+\delta\theta)} $ is minimised when $\vec{\delta\theta}=0$.

Now consider the expression 
\begin{align}
\frac{\partial^2 K(\vec{\theta},\vec{\theta+\delta\theta) }}{\partial \vec{\theta}^2}\Bigg|_{\vec{\delta\theta}=0}
&=\frac{\partial^2}{\partial \vec{\theta^2}}\Bigg (\int   \mathrm{d}\vec{x} \ L(\vec{x}|\vec{\theta})\log \frac{L(\vec{x}|\vec{\theta})}{L(\vec{x}|\vec{\theta+\delta\theta})}\Bigg)\Bigg|_{\vec{\delta\theta}=0}\\
&= \tens{F}(\vec{\theta})\label{eq:Hess} \ ,
\end{align}
from Eq.~(\ref{eq:Fisher}), assuming that integration and differentiation can be interchanged. Thus the Fisher information is the Hessian of the KL divergence given by Eq.~(\ref{eq:KLdefn}) evaluated at ${\vec{\theta+\delta\theta}}$.

Substituting from Eq.~(\ref{eq:Hess}) into Eq.~(\ref{eq:KLapprox}) we finally have a relationship between the KL divergence and the Fisher information 
\begin{align}  \mathrm{KL}(L(\vec{x}|\vec{\theta})\|L(\vec{x}|\vec{\theta+\delta\theta})) 
&=   \frac{1}{2}\vec{\delta\theta}^\intercal\tens{F}(\vec{\theta)}\vec{\delta\theta}  + \mathcal{O} (\|\vec{\delta\theta}\|^3 )\ .\label{eq:KLF}
\end{align}
So, close to the true parameter values, the quantity $\frac{1}{2}\vec{\delta\theta}^\intercal\tens{F}(\vec{\theta)}\vec{\delta\theta}$ is an approximation for the KL divergence. 
Since  mutual information can be defined in terms of the KL divergence between the joint and marginal distributions of two random variables, it follows  that Fisher information and mutual information are related in the limit that $\vec{\delta\theta}$ is sufficiently small in all parameter directions. The KL divergence given by Eq.~(\ref{eq:KLF}) can be considered to be the amount of information lost (or the increase in uncertainty) when $L(\vec{x}|\vec{\theta})$ is approximated by $L(\vec{x}|\vec{\theta+\delta\theta})$. 

\section{Relationships between  compression methods}\label{sect:relations}
Here we summarise some of the key connections and differences between the compression methods described in the main part of the paper.  
\\\\
\noindent\textbf{PCA and MOPED}
Compression based on PCA involves finding the eigenvalues and eigenvectors of the covariance matrix of the data $\tens{C}_\mathrm{d}$.  Fisher-matrix based compression like MOPED involves finding the eigenvalues and eigenvectors of $\tens{J}^\intercal \tens{C}_\mathrm{d}^{-1}\tens{J}$ where $\tens{J}$ is the Jacobian matrix and we assume the covariance does not depend on the parameters. This introduces sensitivity to the parameters.  
Both these eigen problems can be solved by similar standard methods  but their solutions are not directly related.
\\\\
\noindent\textbf{MOPED, score compression and IMNN}
 All three methods use FI as the optimisation criterion. Compression to the score is a non-linear generalisation of MOPED which does not assume a Gaussian likelihood.  IMNN generalises even further by using a neural network to learn  the likelihood from simulations.
\\\\
\noindent\textbf{PCA and CCA}
These can both be expressed as eigen problems but they optimise different criteria. PCA maximises the variance of a single dataset whereas CCA maximises the covariance of two data sets (for data compression, these are the data and the parameters). So the methods are algebraically similar but not generally equivalent. 
\\\\
\noindent\textbf{CCA and VMIM}
By construction VMIM maximises the MI between the data and the parameters. In general CCA does not.  However if the likelihood is Gaussian and the relationship between data and parameters is linear CCA does maximise MI and can be viewed as a restricted version of VMIM.
\\\\
\noindent\textbf{VMIM, GNLL and NN-MSE}
If the estimated posterior in VMIM is replaced by a Gaussian distribution  then it is essentially equivalent to using a GNLL loss function. If additionally the variance of the Gaussian is assumed to be fixed  (just the mean varies with the parameters), then GNLL is equivalent to using an MSE loss function weighted by the covariance of the parameters. If the covariance is the identity then GNLL is equivalent to minimising the MSE of the predicted parameters, which is NN-MSE.

\section{What  Karhunen--Lo\`eve methods are}\label{sect:Karhunen}

\cite{tegmark1997karhunen} referred to compression based on maximising retained Fisher information as    Karhunen--Lo\`eve, introducing this term to cosmology for the first time. This is slightly confusing as the term was originally used differently, and still is in other fields like signal processing.

The \lq  classical\rq \   Karhunen--Lo\`eve  transform (or decomposition) was first developed in the mid-20th century in the context of stochastic processes.   The formalism was developed by several people including Karhunen \citep{karhunen1947lineare},  Lo\`eve \citep{loeve1948functions}, and Hotelling \citep{hotelling1933analysis}.  Their aim was to find a linear transformation of the data in terms of orthogonal basis functions which minimises the mean square error of the reconstructed  data.  It can be shown that the best such transformation is in terms of eigenvectors of the covariance matrix of the data. Thus it is essentially equivalent to PCA. 

As discussed in Sect.~\ref{sect:FI}, \cite{tegmark1997karhunen} changed the problem to that of maximising  Fisher information rather than minimising the MSE of the reconstructed data, but still referred to it as  Karhunen--Lo\`eve. The formalism in their paper is not about optimising reconstruction of the data but about maximising sensitivity to the parameters.  Despite being given the same name, these are distinct optimisation problems.

\section{CCA and mutual information of two Gaussian distributions}\label{sect:GaussianMI}
In this appendix we show that if two distributions are both Gaussian, then CCA maximises their mutual information.

Let $P(\vec{x})$ be a multivariate Gaussian with dimension $n$, and assume the mean of the distribution is zero (if not, we can subtract the mean), and its variance is $\tens{C}_x$. Then by definition
\begin{align}
    P(\vec{x}) &= (2\pi)^{-n/2}(\mathrm{det}(\tens{C}_x))^{-1/2}\exp[-\frac{1}{2}\vec{x}^{\intercal} \tens{C}_x^{-1}\vec{x}] \ .
\end{align}

From Eq.~(\ref{eq:Hx}), the information entropy of $\vec{x}$ is
\begin{align}
    H(\vec{x}) &= -\mathbb{E}\big[\ln((2\pi)^{-n/2}(\mathrm{det}(\tens{C}_x))^{-1/2}\exp(-\frac{1}{2}\vec{x}^{\intercal} \tens{C}_x^{-1}\vec{x}))\big] \notag \\
    &=-\mathbb{E}\big[-\frac{n}{2}\ln(2\pi)-\frac{1}{2}\ln(\mathrm{det}(\tens{C}_x)) -\frac{1}{2}\vec{x}^{\intercal} \tens{C}_x^{-1}\vec{x}\big] \notag \\
    &=\frac{n}{2}\ln(2\pi) +\frac{1}{2}\ln(\mathrm{det}(\tens{C}_x)) +\frac{n}{2} \notag \\
    &= \frac{n}{2}\big[\ln(2\pi) +1\big] +\frac{1}{2}\ln(\mathrm{det}(\tens{C}_x))\label{eq:Gaussian_entropy} \ .
\end{align}
To get to this result we need \mbox{$\mathbb{E}\big[\vec{x}^{\intercal} \tens{C}_x^{-1}\vec{x}]$}.
This can be found by
\begin{align}
  \mathbb{E}\big[\vec{x}^{\intercal} \tens{C}_x^{-1}\vec{x}] &= \mathbb{E}\big[\mathrm{tr}(\vec{x}^{\intercal} \tens{C}_x^{-1}\vec{x})] \notag \\
  &= \mathbb{E}\big[\mathrm{tr}(\tens{C}_x^{-1}\vec{x}^{\intercal} \vec{x})] \notag \\
  &= \mathrm{tr}(\tens{C}_x^{-1}\mathbb{E}\big[\vec{x}^{\intercal} \vec{x}] )\notag \\
  &= \mathrm{tr}(\tens{C}_x^{-1}\tens{C}_x) \notag \\ 
  &= n \ . 
\end{align}
If $\vec{x}$  and $\vec{y}$ are two Gaussian distributions with dimensions $n$ and $m$, respectively, then 
their joint distribution is also Gaussian, with dimension $(n+m)$. 
From Eq.~(\ref{eq:MI}) their mutual information is defined as
\begin{align}
    I(\vec{x};\vec{y}) &= H(\vec{x}) + H(\vec{y}) - H(\vec{x},\vec{y}) \ .
\end{align}
So, using Eq.~(\ref{eq:Gaussian_entropy}),
\begin{align}
    I(\vec{x};\vec{y}) 
    &=\frac{n}{2}\big[\ln(2\pi) +1\big] +\frac{1}{2}\ln(\mathrm{det}(\tens{C}_x))\notag \\
    &\quad\quad\quad+ \frac{m}{2}\big[\ln(2\pi) +1\big] +\frac{1}{2}\ln(\mathrm{det}(\tens{C}_y)) \notag \\
    &\quad\quad\quad-\frac{(n+m)}{2}\big[\ln(2\pi) +1\big] +\frac{1}{2}\ln(\mathrm{det}(\tens{C}_{xy}))\notag \\
    &= \frac{1}{2}\ln\bigg[\frac{\mathrm{det}(\tens{C}_{x})\mathrm{det}(\tens{C}_{y})}{\mathrm{det}(\tens{C}_{xy})}\bigg]\ ,\label{eq:IxyGauss}
\end{align}
where $\tens{C}_{xy}$ is the covariance between $\vec{x}$ and $\vec{y}$.

In the particular case of CCA (Sect.~\ref{sect:CCA}) we require the mutual information between the parameters $\vec{p}$ and the compressed vector $\vec{b}^\intercal\vec{d}$. If both $\vec{p}$ and $\vec{d}$ have Gaussian distributions, we can use Eq.~(\ref{eq:IxyGauss}) to get 

\begin{equation}\label{eq:Ipbd}
    I(\vec{p};\vec{b}^\intercal\vec{d}) = \frac{1}{2}\ln\bigg[\frac{\mathrm{det}(\tens{C}_{\mathrm{p}})\mathrm{det}(\vec{b}^\intercal\tens{C}_{\mathrm{d}}\vec{b})}{\mathrm{det}(\tens{C}^\prime)}\bigg] \ ,
\end{equation}
where, following \cite{park2025dimensionality}, $\tens{C}^\prime$ is the covariance between $\vec{p}$ and $\vec{b}^\intercal\vec{d}$, given by
\begin{equation}
\tens{C}^\prime = \begin{pmatrix}
\tens{C}_{\mathrm{p}} &\tens{C}_{\mathrm{pd}}\vec{b} \\
\vec{b}^\intercal\tens{C}_{\mathrm{dp}} & \vec{b}^\intercal\tens{C}_{\mathrm{d}}\vec{b} 
\end{pmatrix} \ .\\
\end{equation}
$\tens{C}_{\mathrm{d}}$ and $\tens{C}_{\mathrm{p}}$ are the auto-covariances of the data and parameters respectively, $\tens{C}_{\mathrm{pd}}$ is their cross-covariance, and \mbox{$\tens{C}_{\mathrm{dp}}= \tens{C}_{\mathrm{pd}}^\intercal$}.

Then Eq.~(\ref{eq:Ipbd}) can be re-written as:
\begin{align}
     I(\vec{p};\vec{b}^\intercal\vec{d}) 
     &= \frac{1}{2} \ln \Bigg[\frac{\mathrm{det}(\tens{C}_{\mathrm{p}})\vec{b}^\intercal\tens{C}_{\mathrm{d}}\vec{b}}{\mathrm{det}(\tens{C}^\prime)}\bigg]  \notag\\
     &= 
     \frac{1}{2} \ln \Bigg[\frac{\mathrm{det}(\tens{C}_{\mathrm{p}})\vec{b}^\intercal\tens{C}_{\mathrm{d}}\vec{b}}
     {\mathrm{det}(\tens{C}_{\mathrm{p}})({\vec{b}^\intercal\tens{C}_{\mathrm{d}}\vec{b}-\vec{b}^\intercal\tens{C}_{\mathrm{dp}}\tens{C}_{\mathrm{p}}^{-1}}\tens{C}_{\mathrm{pd}}\vec{b})}\Bigg]\notag \\        
     &=\frac{1}{2} \ln \Bigg[\frac{\vec{b}^\intercal\tens{C}_{\mathrm{d}}\vec{b}}{{\vec{b}^\intercal\tens{C}_{\mathrm{d}}\vec{b}-\vec{b}^\intercal\tens{C}_{\mathrm{dp}}\tens{C}_{\mathrm{p}}^{-1}}\tens{C}_{\mathrm{pd}}\vec{b}}\Bigg] \ .
\end{align}

We can require that $\vec{b}^\intercal\tens{C}_{\mathrm{d}}\vec{b} = 1$, so that maximising $ I(\vec{p};\vec{b}^\intercal\vec{d})$ is equivalent to minimising $\vec{b}^\intercal\tens{C}_{\mathrm{dp}}\tens{C}_{\mathrm{p}}^{-1}\tens{C}_{\mathrm{pd}}\vec{b}$ subject to this constraint.
This leads to the Lagrangian
\begin{align}
    \Lambda = \vec{b}^\intercal\tens{C}_{\mathrm{d}}\vec{b} - \rho\vec{b}^\intercal\tens{C}_{\mathrm{dp}}\tens{C}_{\mathrm{p}}^{-1}\tens{C}_{\mathrm{pd}}\vec{b} \ ,
\end{align} 
where $\rho$ is a Lagrangian multiplier.
Differentiating with respect to $\vec{b}$ and equating to zero  gives
\begin{align}
 \tens{C}_{\mathrm{d}}\vec{b} - \rho \tens{C}_{\mathrm{dp}}\tens{C}_{\mathrm{p}}^{-1}\tens{C}_{\mathrm{pd}}\vec{b}=0 \ .
\end{align}
Rearranging this, 
\begin{align}
\tens{C}_{\mathrm{d}}^{-1}\tens{C}_{\mathrm{dp}}\tens{C}_{\mathrm{p}}^{-1}\tens{C}_{\mathrm{pd}}\vec{b} &= \rho \, \vec{b} \ , 
\end{align}
which is identical to Eq.~(\ref{eq:gammasqb}) with $\gamma^2 = \rho$.
Thus when the probability distributions of $\vec{p}$ and $\vec{d}$ are both Gaussian, maximising the correlation of $\vec{p}$ and $\vec{d}$ with CCA is equivalent to maximising their mutual information.

\bsp
\label{lastpage}
\end{document}